\documentclass[aps,prd,amsmath,showpacs, showkeys,onecolumn,floatfix, nofootinbib 
, 11pt]{revtex4-1} 

\usepackage{bm}
\usepackage{graphicx}
\usepackage{hyperref}
\usepackage{slashed}
\usepackage{subfigure}
\usepackage[svgnames]{xcolor}

\begin{document}

\begin{flushright} ADP-11-12/T734 \end{flushright}

\title{Monte Carlo Simulations of Hadronic Fragmentation Functions using NJL-Jet Model}
\author{Hrayr H.~Matevosyan}
\author{Anthony W. Thomas}
\affiliation{CSSM, School of Chemistry and Physics, \\
University of Adelaide, Adelaide SA 5005, Australia\\
http://www.physics.adelaide.edu.au/cssm
}
\author{Wolfgang Bentz}
\affiliation{Department of Physics, School of Science,\\  Tokai University, Hiratsuka-shi, Kanagawa 259-1292, Japan\\
http://www.sp.u-tokai.ac.jp/
}
\date{\today}                                           

\begin{abstract}
The recently developed Nambu-Jona-Lasinio (NJL) - jet model is used as an
effective chiral quark theory to calculate the quark fragmentation functions
to pions, kaons, nucleons, and antinucleons. The effects of the vector mesons
$\rho$, $K^*$, and $\phi$ on the production of secondary pions and kaons are included.
The fragmentation processes to nucleons and 
antinucleons are described by using the quark-diquark picture, which has been shown to give
a reasonable description of quark distribution functions. We incorporate
effects of next-to-leading order in the $Q^2$ evolution, and compare
our results with the empirical fragmentation functions.
\end{abstract}

\maketitle

\section{Introduction}
\label{SEC_INTRO} 

The understanding of quark fragmentation functions has been rapidly evolving 
in recent years~\cite{Accardi:2009qv}. Based on precise data on inclusive hadron production
in hard scattering processes~\cite{Buskulic:1994ft,Abreu:1998vq,Abbiendi:1999ry,Aihara:1986mv,Abe:1998zs,Adler:2003pb,Arsene:2007jd,Abelev:2006cs}, empirical fragmentation functions have
been extracted, and their behavior under scale evolutions has been 
investigated in detail~\cite{deFlorian:2007aj, deFlorian:2007hc,Hirai:2007cx}. It is reasonable to expect that in the near future
an understanding of the fragmentation functions will be attained which
is as precise as the present knowledge on quark distribution functions.
Very interesting prospects for the study of fragmentation processes
were opened by the HERMES~\cite{Airapetian:2007vu} and JLab~\cite{Avakian:2010ae, Avakian:2005ps,Mkrtchyan:2007sr,Brooks:2009xg} semi-inclusive 
measurements on nuclear
targets, which will be followed up by the planned Electron-Ion Collider~\cite{Brooks:2010rz,Thomas:2009ei}.  
  
These exciting developments present a challenge for effective theories of
QCD. Because both the distribution and fragmentation functions are basically
nonperturbative quantities, effective quark theories are powerful tools for
theoretical studies. 
In particular, the Nambu-Jona-Lasinio (NJL) model~\cite{Nambu:1961tp,Nambu:1961fr}, which was very successful in the description of quark distribution functions~\cite{Cloet:2005pp}, has been combined recently
with the ideas of the jet model of Field and Feynman~\cite{Field:1977fa} to give a realistic
description of quark fragmentation functions~\cite{Ito:2009zc}. In recent work we have
applied this NJL-jet model to the fragmentation functions for pions and
kaons, and compared the results to the empirical functions~\cite{Matevosyan:2010hh}. The main
advantage of this model, which is based on the multiplicative ansatz 
(product ansatz) of Field and Feynman, is that the fragmentation functions automatically
satisfy the important momentum and isospin sum rules without introducing
any new parameters into the theory. Ultimately, the NJL-jet model will provide a consistent framework to describe
semi-inclusive deep inelastic lepton-hadron scattering, as well as allowing 
predictions for nuclear targets.

The purpose of the present paper is to extend the NJL-jet calculations
of quark fragmentation functions in the following three directions: 
First, we include the effects of secondary pions and kaons, which come from the decay of
intermediate $\rho$, $K^*$, and $\phi$ mesons. We also include the strong decays of the
vector mesons to $\pi$  and $K$ two-particle final states. 
Our aim is to investigate, in particular, the role of the $\rho$ meson for the
softening of the pion momentum distribution. The effects of three-particle decays,
like the $\omega$ meson, will be left for future work, as those processes require a treatment
of nontrivial phase space factors that cannot be evaluated analytically,
which goes beyond the usual convolution formalism. 

Second, we will include the fragmentation processes
to nucleons and antinucleons.  For this purpose, we will use the
splitting functions obtained in the quark-diquark description of
baryons~\cite{Mineo:1999eq}, taking into account only the effects of the scalar diquark as a first step in extending the model.  The need to include the axial-vector diquark to fully describe the nucleon structure and fragmentation functions has been demonstrated in the earlier work~\cite{Jakob:1997wg,Bacchetta:2003rz,Bacchetta:2008af}, thus this remains a priority for the future developments of the model.  The NJL description of nucleons as
bound states of a quark and a scalar diquark has already been applied to fragmentation functions in previous work~\cite{Kitagawa:2001ig,Yang:2002gh}. In the present paper we will, however, go beyond those earlier attempts
by including the elementary fragmentations to nucleons and antinucleons
in the cascadelike processes also including pions, kaons and vector mesons.

Third, we use the Monte Carlo (MC) method to solve for the fragmentation functions within the quark-cascade model as opposed to solving integral equations in previous work. We demonstrate the viability of this approach to replace the integral equations by providing very similar order of precision in determining the fragmentation functions, at the same time allowing to relax the approximations necessary for formulating the integral equations and easily incorporating the resonance decays i nto the model.

 This paper is organized as follows: In Sec. \ref{SEC_SPLITT},  we present the calculations of the vector meson elementary fragmentation functions and elementary fragmentation functions of a quark to nucleon antinucleon pair.  In Sec. \ref{SEC_MC}, we describe the Monte Carlo method for calculating fragmentation functions and include the vector meson decays. In Sec. \ref{SEC_RES}, we present the resulting solutions for the fragmentation functions. Section \ref{SEC_CONC} contains some concluding remarks and a future outlook for extending the model. 
  
\section{Elementary Fragmentation Functions}
\label{SEC_SPLITT}
In this article, we extend our previous work of Ref.~\cite{Matevosyan:2010hh} using the same notation and model parameters. In the current section, we evaluate the "elementary" fragmentation functions of quarks to hadrons as a "one-step" process in the NJL model using light-cone (LC) coordinates\footnote{ We use the following LC convention for Lorentz 4-vectors $(a^+,a^-,\mathbf{a_\perp})$, $a^\pm=\frac{1}{\sqrt{2}}(a^0\pm a^3)$ and $\mathbf{a_\perp} = (a^1,a^2)$. }. The NJL model we use includes only four-point quark interaction in the Lagrangian, with up, down, and strange quarks and no additional free parameters (see, e.g. Refs.~\cite{Kato:1993zw,Klimt:1989pm,Klevansky:1992qe} for detailed reviews of the NJL model). We employ Lepage-Brodsky (LB) ``invariant mass'' cut-off regularization for loop integrals (see Refs.~\cite{Matevosyan:2010hh} for a detailed description as applied to the NJL-jet model), except when calculating meson-quark couplings as discussed further in this section, and use our previous values for the constituent quark masses for light and strange quarks $M_u=300~\rm{MeV}$,  $M_s=537~\rm{MeV}$. 
  
\subsection{Pseudoscalar Meson Splitting Functions}
\label{SEC_PS}

\begin{figure}[h]
\includegraphics[width=0.4\textwidth]{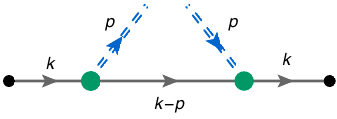}
\caption{Quark splitting function for mesons.}
\label{PLOT_FRAG_QUARK}
\end{figure}

The elementary fragmentation function of quark $q$ emitting a meson $m$ carrying light-cone momentum fraction $z$ is depicted in Fig.~\ref{PLOT_FRAG_QUARK}. For completeness of the description we present the results for the pseudoscalar mesons derived in~\cite{Matevosyan:2010hh}, leaving out the details. In the frame where the fragmenting quark has  $\mathbf{k_\perp}=0$, but nonzero transverse momentum component $\mathbf{k}_T=-\mathbf{p_\perp}/z$ with respect to the direction of the produced hadron, the relation for the elementary fragmentation function is
\begin{align}
\label{EQ_QUARK_FRAG}
\nonumber
d_{q}^{m}(z)= - \frac{C_q^m}{2}  g_{mqQ}^{2} \frac{z}{2} \int \frac{d^{4}k}{(2\pi)^{4}} Tr[S_{1}(k)\gamma^{+}S_{1}(k)\gamma_{5} (\slashed{k}-\slashed{p}+M_{2}) \gamma_{5}]&\\ 
 \times \delta(k_{-} - p_{-}/z) 2 \pi \delta( (p-k)^{2} -M_{2}^{2} )
&\\ 
= \frac{C_q^m}{2}  g_{mqQ}^{2} z \int \frac{ d^{2} \mathbf{p}_{\perp}}{(2\pi)^{3}} \frac{p_{\perp}^{2}+((z-1)M_{1}+M_{2})^{2}} {(p_{\perp}^{2}+z(z-1)M_{1}^{2}+zM_{2}^{2}+(1-z)m_{m}^{2})^{2}}.&
\end{align}

Here $\mathrm{Tr}$ denotes the Dirac trace and the subscripts on the quark propagators denote quarks of different flavor - also indicated by $q$ and $Q$, where the meson of type $m$ under consideration has the flavor structure $m=q\overline{Q}$. The $C_q^m$ is the corresponding flavor factor given in the Table~\ref{TB_FLAVOR_FACTORS}. The pseudoscalar meson-quark coupling constant, $g_{mqQ}$, is determined from the residue at the pole in the quark-antiquark t-matrix at the mass of the meson under consideration.

In LB regularization the cut-off in the transverse momentum, $P_\perp^2$, is given by 
\begin{eqnarray}
\label{EQ_QUARK_FRAG_LB_PPERP}
P_\perp^2= z(1-z)\left(\sqrt{\Lambda_{3}^2 +m_m^2}+  \sqrt{\Lambda_{3}^2 +M_2^2}\right)^2 - (1-z)m_m^2 -z M_2^2,
 \end{eqnarray}
where $\Lambda_3= 0.67 ~{\rm GeV}$ denotes the 3-momentum cutoff obtained in Ref.~\cite{Bentz:1999gx}, yielding the following values for the meson-quark couplings:
\begin{align}
\label{EQ_PS_Q_COUPLING}
g_{\pi qQ}=3.15 ,\ \ 
g_{K qQ}=3.39 .
\end{align}

 A consequence of LB regularization is a limited range for $z$ in corresponding regularized functions, $0<z_{min}\leq z\leq z_{max}<1$, where $z_{min}$ and $z_{max}$ are determined by imposing the condition $P_{\perp}^2\geq 0$ in Eq.~(\ref{EQ_QUARK_FRAG_LB_PPERP}). These range limitations depend on the masses of hadrons and quarks involved. For example, the $z$ limits are very close to endpoints ($z=0$ and $z=1$) for splitting functions to pions, but are quite far from them for heavier hadrons like nucleons or $\phi$ meson.

\subsection{Vector Meson Splitting Functions}
\label{SEC_VM}

 In the experimental measurements of the quark hadronization process, one usually directly detects only the relatively long lived particles: pions, nucleons, and sometimes kaons. These particles can come either as direct emissions of the fragmenting partons or as decay products of hadronic resonances. We note that in the current article we will only consider the strong decays of the hadrons. The inclusion of the vector mesons in the NJL-jet model is important, since it has a two-fold effect on the description of the previously calculated pion and kaon fragmentation functions.
  First, the availability of the new emission channels decreases the normalization of the direct quark splitting functions to pions and kaons. In fact, in the early models for the quark fragmentation, it was assumed that the ratio of the vector to pseudoscalar meson fragmentations should follow from the simple spin state counting rule, $3:1$. Later, it was shown experimentally that this vector meson ratio is significantly smaller, especially in the light quark sector (see, for example, Ref.~\cite{Chliapnikov:1999qi}), and thus has been attributed to dynamic effects like the masses of the considered mesons having a strong influence on the ratio. (The above mentioned rule holds only for those mesons containing the heaviest quarks, where the differences in masses between pseudoscalar and vector mesons are negligible.)
  Second, the strong decays of the vector mesons produce pseudoscalar mesons with a $z$ distribution that is distinguishable from the ones emitted directly in the quark fragmentation process, thus modifying the final $z$ dependence of the fragmentation functions.

The basic NJL interaction Lagrangian relevant for the vector meson channels has the form $-G_v \left( \bar{\psi}\gamma^\mu \lambda_\alpha \psi \right)^2$, where $\alpha = 0, 1, . . .8$ with $\lambda_0 = \sqrt{2/3}\ \mathbf{1}$. If the 4-Fermi coupling constant $G_v$ is chosen to reproduce the mass of the $\rho$ meson as the pole of the $q\bar{q}$ t-matrix, the calculated masses of $K^*$ and $\phi$ mesons agree well with the experimental values (see Appendix~\ref{SUB_SEC_VM_COUPL}). Using the vector meson - quark vertex $g_{mqQ}\gamma^\mu \lambda_\alpha$ (versus $ \imath g_{mqQ}\gamma^5 \lambda_\alpha$ for pseudoscalar mesons), the elementary fragmentation function can be written as
\begin{align}
\label{EQ_QUARK_FRAG_V}
\nonumber
d_{q}^{m}(z)=  \sum \limits_{\lambda=\mp1, 0} \frac{C_{q}^m}{2}  g_{mqQ}^{2} \frac{z}{2}
\int \frac{d^{4}k}{(2\pi)^{4}}
\mathrm{Tr}[S_{1}(k)\gamma^{+}S_{1}(k)\gamma_{\mu} (\slashed{k}-\slashed{p}+M_{2}) \gamma_{\nu}]  \epsilon^\mu(\lambda, p) \epsilon^{*\nu}&(\lambda, p)\\
 \times \delta(k_{-} - p_{-}/z) 2\pi  \delta( (k-p)^{2}& -M_{2}^{2} ),
\end{align}
where the sum is over the polarization\footnote{Since in this paper we consider only spin-independent fragmentation functions, we include the spin degeneracy factor $2s_b + 1$ into the definition of the fragmentation function $d_a^b(z)$, so as to facilitate the comparison to the empirical parametrizations~\cite{Hirai:2007cx,deFlorian:2007aj}. (We note that the operator definition used in Ref.~\cite{Ito:2009zc} does not include this degeneracy factor.)} of the vector meson $m$ and $Tr$ is over Dirac indices. The details of the calculation of the $Tr$ and integration over plus and minus components of the momentum are given in Appendix~\ref{APP_VM_SPLITT}. Here we simply present the resulting expression:
 \begin{eqnarray}
\label{EQ_QUARK_FRAG_F1}
d_{q}^{m}(z)
=\frac{C_{q}^m}{2}  g_{mqQ}^{2} z 
\int \frac{ d^{2} \mathbf{p}_\perp}{(2\pi)^{3}} &
 \frac{
 2  \left(p_\perp^2 +((1-z)M_1-M_2)^2\right)
+\frac{1}{z^2 m_m^2}\left( \left(p_\perp^2 - z^2 M_1 M_2+(1-z) m_m^2\right)^2+p_\perp^2 z^2 \left(M_1+M_2\right)^2  \right)
 }
{ \left (p_\perp^2 + z(z-1)M_1^2+ z M_2^2 + (1-z)m_m^2  \right)^2}.& \nonumber \\
\end{eqnarray}

 The vector meson-quark couplings, $g_{mqQ}$, are determined from the residue at the pole in the quark-antiquark t-matrix at the mass of the meson under consideration. Here we encounter a deficiency in LB regularization scheme, where the mesons with mass larger than the sum of constituent quark's masses are not bound. This problem is usually circumvented \cite{Ebert:1996vx} by choosing a large enough constituent quark mass accounting for the masses of all the considered hadrons. Here to keep consistency with our previous calculations, we choose to use a different approach: we will use the same constituent quark masses as in the previous NJL-jet model calculations, but in order to assess the vector meson-quark couplings we will use the proper-time (PT) regularization scheme of Refs.~\cite{Ebert:1996vx, Hellstern:1997nv, Bentz:2001vc} that mimics confinement and eliminates the unphysical decay thresholds. The details of these calculations are presented in Appendix~\ref{APP_QM_COUPLING}, where we compare the pseudoscalar meson-quark couplings calculated both in LB and PT regularization schemes and ensure that they are practically the same. Also, the strange constituent quark mass determined from the experimentally measured kaon mass is practically the same in both regularization schemes. It is therefore reasonable to use the PT scheme for the calculation of the vector meson- quark couplings in the present model, while keeping the LB scheme for the regularization of the elementary fragmentation functions.
 
\subsection{$N$ and $\overline{N}$ fragmentation channels}
\label{SEC_NNB}

\begin{figure}[phtb]
\centering 
\subfigure[] {
\includegraphics[width=0.48\textwidth]{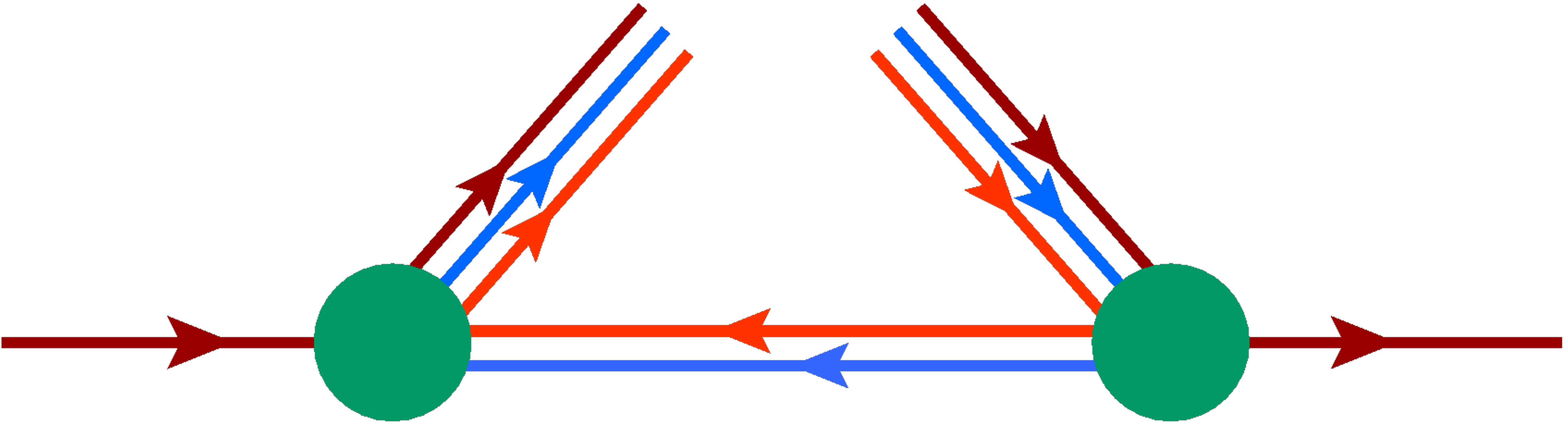}}
\hspace{0.1cm} 
\subfigure[] {
\includegraphics[width=0.48\textwidth]{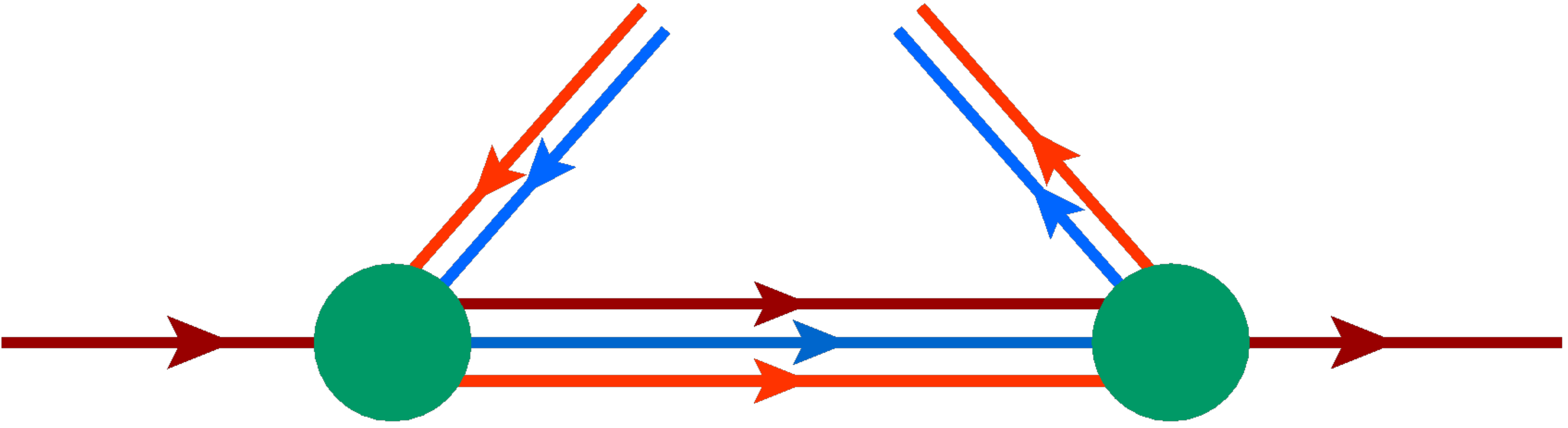}
}
\vspace{0.1cm} 
\subfigure[] {
\includegraphics[width=0.98\textwidth]{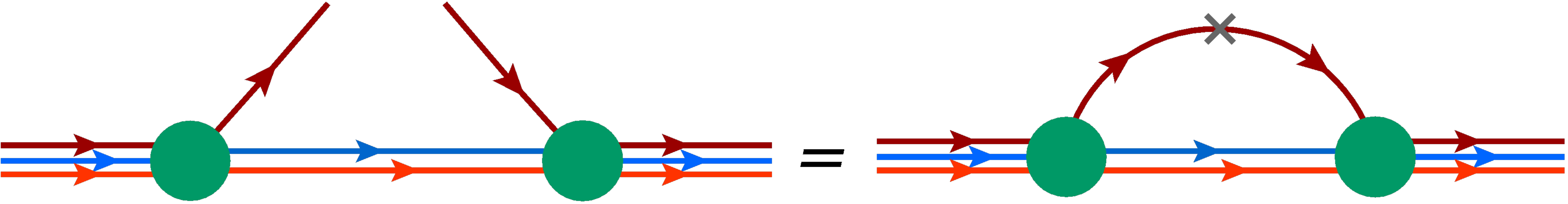}
}
\caption{Two contributions to the quark fragmentation function to a nucleon and antidiquark are depicted on Figs.  (a) and (b),  and the quark distribution function in nucleon $f_q^N(x)$ is depicted in  (c) as a cut diagram (left) and as the equivalent Feynman diagram (right).}
\label{PLOT_FRAG_Q_N}
\end{figure}

\begin{figure}[phtb]
\centering 
\subfigure[] {
\includegraphics[width=0.48\textwidth]{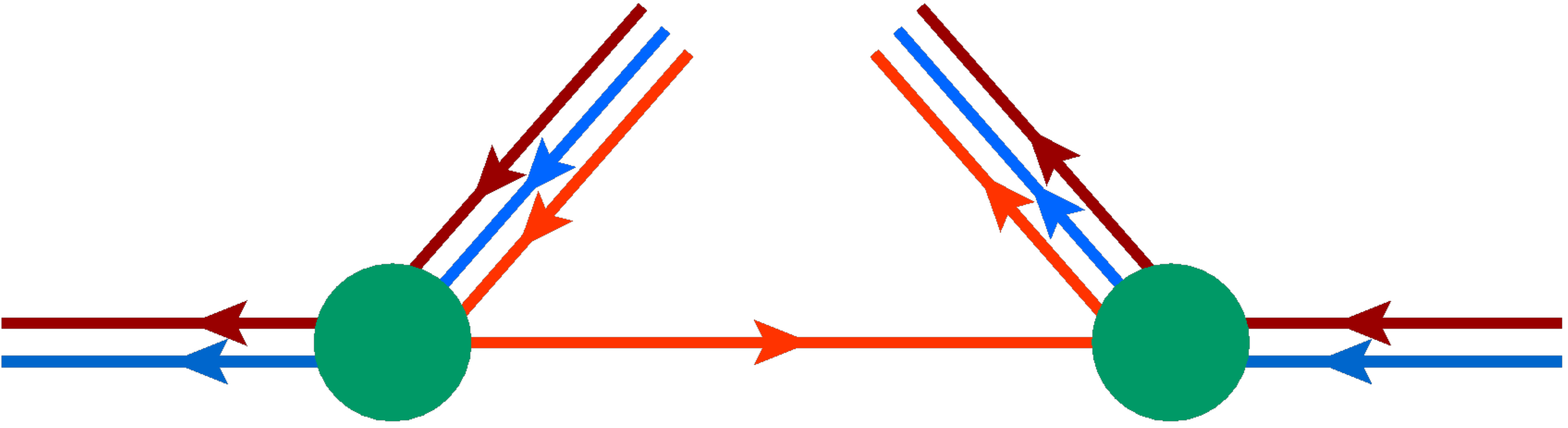}}
\hspace{0.1cm} 
\subfigure[] {
\includegraphics[width=0.48\textwidth]{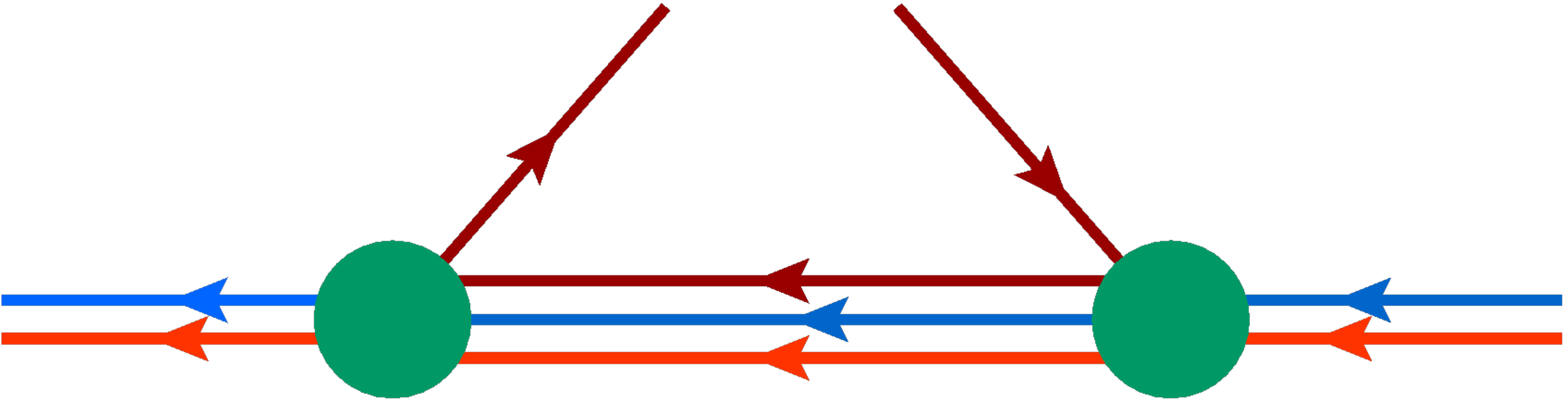}
}
\caption{Two contributions to the antidiquark fragmentation function to an antinucleon and a quark are depicted on Figs.  (a) and (b).}
\label{PLOT_FRAG_DQB_N}
\end{figure}

In this section, we consider the elementary splitting functions, which involve 
nucleons and \\ antinucleons, i.e., $q \rightarrow N\overline{D}$ and the subsequent 
process $\overline{D} \rightarrow \overline{N} q$. Here $q=u,\,d$ denotes the
nonstrange quark, $N=n,\,p$ denotes
the nucleon, and $D$ denotes a scalar
diquark ($J=T=0, \, \overline{3}_c$). The first process leads to the
elementary fragmentation functions $d_q^N(z)$, represented as a cut-
diagram in Fig.~\ref{PLOT_FRAG_Q_N}a, and $d_q^{\overline{D}}(z)$, which is shown in Fig.~\ref{PLOT_FRAG_Q_N}b.
We recall that for a general process $a \rightarrow b c$, the
fragmentation function $d_a^b(z)$ corresponds to the probability distribution
of the LC momentum carried by $b$ relative to $a$, and includes a sum over the 
spin-color quantum numbers of the spectator $c$. Therefore, the two
fragmentation functions of Fig.~\ref{PLOT_FRAG_Q_N} are simply related by
\begin{eqnarray}
d_q^{\overline D}(z) = \frac{1}{3} d_q^N(1-z) \,.   \label{rel1}
\end{eqnarray}
The second process $\overline{D} \rightarrow \overline{N} q$ leads to the
elementary fragmentation functions $d_{\overline D}^{\overline N}(z)$, shown in 
Fig.~\ref{PLOT_FRAG_DQB_N}a, and $d_{\overline D}^q(z)$ of Fig.~\ref{PLOT_FRAG_DQB_N}b. The relation for this case is
\begin{eqnarray}
d_{\overline D}^q(z) = \frac{1}{3} d_{\overline D}^{\overline N}(1-z) \,. \label{rel2}
\end{eqnarray}
To evaluate the functions $d_q^N(z)$ and $d_{\overline D}^{\overline N}(z)$, 
one can either directly consider the
cut diagrams of Figs.~\ref{PLOT_FRAG_Q_N}a and~\ref{PLOT_FRAG_DQB_N}a, or one can make use of the fact that all
elementary NJL fragmentation functions are formally related to the more familiar 
distribution functions $f(x)$ by crossing and charge conjugation, which is expressed
by the relation\footnote{We emphasize that this so called ``Drell-Levy-Yan
relation'' can be used only to obtain the expressions for the {\em elementary}
{\em unregularized} fragmentation functions from the distribution functions.} \cite{Drell:1969jm,Blumlein:2000wh}
\begin{eqnarray}
d_a^b(z) = \left(-1 \right)^{2(s_a+s_b) + 1} (2s_b+1)\frac{z}{\gamma_a} 
f_a^b\left(x=\frac{1}{z}\right) \,,  \label{dly}
\end{eqnarray}
where $\gamma_a$ is the spin-color degeneracy of $a$.
Therefore, in order to obtain $d_q^N(z)$, we can use the well-known 
expression for the quark distribution function inside a nucleon, where the 
nucleon is
described as a bound state of a quark (mass $M$) and a scalar diquark\footnote{The basic NJL interaction Lagrangian in the scalar diquark (D) channel	has	the	form	$G_s \left(\bar{\psi} \gamma_5 C \tau_2 \beta^A \bar{\psi}^T\right) \left(\psi^T  C^{-1} \gamma_5 \tau_2 \beta^A \psi \right)$, where	$\beta^A= \sqrt{3/2}\ \lambda_A\ (A = 2, 5, 7)$ and $C = i\gamma_2 \gamma_0$. The diquark mass $M_D$ is calculated as the pole of the $qq$ t-matrix, and the nucleon mass as the pole of the $qD$ t-matrix~\cite{Bentz:2001vc}. The 4-Fermi coupling constant $G_s$ is chosen so as to reproduce the experimental nucleon mass.}
(mass $M_D$). This
distribution function $f_q^N(x)$ is shown in Fig.~\ref{PLOT_FRAG_Q_N}c. It is easily evaluated
by using the form of the nucleon vertex function \cite{Mineo:2003vc}
\begin{eqnarray}
\Gamma_N(p) = \sqrt{- Z_N \frac{M_N}{p_-}} \, u_N(p) \,, 
\label{vert}
\end{eqnarray}
where the Dirac spinors are normalized as $\overline{u}_N u_N = 1$, and the normalization
factor $Z_N$ is given by
\begin{eqnarray}
{Z}_N = \left(- \frac{\partial {\Pi}_N}{\partial \slashed{p}}
\right)^{-1}_{\slashed{p}=M_N}\,,   \label{zn}
\end{eqnarray}
with the quark-diquark bubble graph expressed by the Feynman
propagators of the quark and the diquark:
\begin{eqnarray}
{\Pi}_N(p) = i \int \frac{{\rm d}^4 k}{(2\pi)^4} S_F(k) \Delta_s(p-k) \,,
\label{qdbub}
\end{eqnarray} 
 which yields for the normalization factor
 \begin{eqnarray}
Z_N^{-1} &=& \frac{1}{2} \int_0^1 {\rm d}x
\int \frac{{\rm d}^2 \mathbf{p}_{\perp}}{(2\pi)^3}
\frac{(1-x)\left(p_{\perp}^2+(M_Nx+M)^2 \right)}{\left[
M_N^2 x(1-x) - p_{\perp}^2 - M^2 - x(M_D^2-M^2)\right]^2}.
\end{eqnarray}

We obtain from the Feynman diagram of Fig.~\ref{PLOT_FRAG_Q_N}c
\begin{eqnarray}
f_q^N(x) &=& i M_N {Z}_N {\overline u}_N \int \frac{{\rm d}^4 k}
{(2\pi)^4} \left(S_F(k) \gamma^+ S_F(k) \right) \Delta_s(p-k) 
\delta(k_- - p_- x) u_N  \nonumber \\
&=& \frac{1}{2} {Z}_N (1-x) \int \frac{{\rm d}^2 \mathbf{k}_T}
{(2\pi)^3}  \frac{{k}_T^2 + (M_N x + M)^2}{
\left[{ k}_T^2 + M^2(1-x) + M_D^2 x - M_N^2 x (1-x)\right]^2}\,.
\label{fqn}
\end{eqnarray}
Here, $(q,N) = (u,p)$ or $(d,n)$, while for the other flavor combinations the 
distribution function
vanishes. The relation (\ref{dly}) then gives
\begin{eqnarray}
d_q^N(z) = \frac{1}{6} {Z}_N (1-z)
\int \frac{{\rm d}^2 \mathbf{p}_{\perp}} {(2\pi)^3}
\frac{{ p}_{\perp}^2 + (M_N + M z)^2}
{\left[{ p}_{\perp}^2 - M^2 z(1-z) + M_D^2 z + M_N^2 (1-z)\right]^2}\,.
\label{dqn}
\end{eqnarray}
The function $d_q^{\overline D}(z)$ can then be obtained from (\ref{rel1}),
which completes the evaluation of the fragmentation functions of Figs.~\ref{PLOT_FRAG_Q_N}a and~\ref{PLOT_FRAG_Q_N}b.

For the fragmentation function $d_{\overline D}^{\overline N}(z)$ of
Fig.~\ref{PLOT_FRAG_DQB_N}a, we note that charge conjugation invariance gives
$d_{\overline D}^{\overline N}(z) = d_{D}^{N}(z)$, where the latter function is
related to the diquark distribution function inside the nucleon ($f_D^N(x)$)
by the crossing relation (\ref{dly}). The function $f_D^N(x)$ in turn is simply
obtained from the quark distribution function (\ref{fqn}) by replacing 
$x \rightarrow 1-x$. Summarizing, we obtain
\begin{eqnarray}
d_{\overline D}^{\overline N}(z) &=& z \frac{2}{3}  f_q^N(1-x) |_{x=1/z} \nonumber \\  
&=& \frac{1}{3} {Z}_N \int \frac{{\rm d}^2 \mathbf{p}_{\perp}}
{(2\pi)^3} \frac{{ p}_{\perp}^2 + \left(M z - M_N(1-z)\right)^2}
{\left[{ p}_{\perp}^2 + M^2 z - M_D^2 z (1-z) + M_N^2 (1-z)\right]^2}\,,
\label{ddn}
\end{eqnarray} 
for both cases $N=p$ or $n$. 
The remaining function $d_{\overline D}^q(z)$ is then obtained
from the relation (\ref{rel2}). This concludes the evaluation of the fragmentation
functions of Figs.~\ref{PLOT_FRAG_DQB_N}a,~\ref{PLOT_FRAG_DQB_N}b.

 Here, we note that another possible channel for a nucleon emission within NJL formalism, $q \rightarrow D\overline{q}$ and subsequent  $ D \rightarrow N\overline{q}$ is not included in the current version of the model, as numerically the norm for the splitting function of this channel is 2 orders of magnitude smaller than the norm of the splitting function for the channel described above, thus proving insignificant.
  
 \section{Monte Carlo Approach to Calculating the Fragmentation Functions}
\label{SEC_MC}
 
 \subsection{Monte Carlo Simulations as an Alternative to the Integral Equation Method}
 \label{SEC_MC_SIMS}
  
   The Monte Carlo method for describing quark fragmentation process using most notably the Lund string model \cite{Andersson:1983ia} has been long employed to successfully describe the hadronization process \cite{Sjostrand:1982fn} and has been implemented in very sophisticated event generator frameworks like PYTHIA~\cite{Sjostrand:2007gs}. These frameworks have been refined, extended, and tuned over several decades to cater for the needs of the experimental data analysis. Our purpose here is to develop a standalone MC simulation software that has the specific purpose of implementing the quark-cascade description of the hadronization process with quark splitting functions supplied from an effective quark model, particularly NJL-jet model in the current article, and vector meson decays to pseudoscalar mesons. This allows for flexibility and simplicity of the software platform, making it readily accessible for further development.
 
 Previously, in the NJL-jet model the fragmentation functions were obtained as solutions of a set of coupled integral equations (\cite{Ito:2009zc,Matevosyan:2010hh}),
\begin{equation} 
\label{EQ_FRAG_PROB}
D^{h}_{q}(z)dz=\hat{d}^{h}_{q}(z)dz+\sum_{Q}\int^{1}_{z}\hat{d}^{Q}_{q}(y) dy  \; D^{h}_{Q}\left(\frac{z}{y}\right) \frac{dz}{y},
\end{equation}
where  $D^h_q(z)$ denotes the fragmentation function of quark $q$ to hadron $h$ carrying light-cone momentum fraction $z$, and  $\hat{d}_{q}^{h}(z)$, $\hat{d}_{q}^{Q}(z)$  are the elementary fragmentation functions for the process $q\to h Q$, normalized as  $\sum_{h}\int \hat{d}_{q}^{h}(z) dz=\sum_{Q}\int \hat{d}_{q}^{Q}(z) dz=1$, thus allowing an interpretation as the probability of an elementary process.  The sum on the right-hand side is over all possible intermediate states in the quark cascade ( in our model that includes all the active flavors of quarks and scalar antidiquarks).  In formulating the integral equations, one assumes that the quark has infinite momentum and produces an infinite number of hadrons, which physically corresponds to the Bjorken limit. 
 
 We propose to calculate the fragmentation functions using Monte Carlo simulations akin to the method described in Ref.~\cite{Ritter:1979mk} using the probabilistic interpretation: $D_q^h(z)\ dz $  is the probability to emit a hadron $h$ carrying the light-cone momentum fraction $z$ to $z + dz$ of initial quark $q$ in a quark-jet picture. The quark goes through a cascade of hadron emissions, where at every emission vertex we choose the type of emitted hadron $h$ and its light-cone momentum fraction $z$ (of the fragmenting quark) by randomly sampling the corresponding probability densities of the elementary fragmentations, $\hat{d}_q^h(z)$, that are calculated within the NJL model (in general these can be calculated in any effective quark model). We keep track of the flavor and the light-cone momentum fraction of the initial quark left to the remaining  quark, also recording the type and light-cone momentum fraction of the initial quark transferred to the emitted hadron in each elementary fragmentation process. We stop the fragmentation chain after the quark has emitted a predefined number of hadrons, $N_{Links}$. (Alternatively, one could stop the chain after the remnant quark in the cascade  has less than a given fraction of the initial quark's light-cone momentum, $z_{Min}$.) We repeat the calculation $N_{Sims}$ times with the same initial quark flavor $q$, until we have sufficient statistics for the emitted hadrons. We extract the fragmentation functions by calculating the average number of hadrons of type $h$ with light-cone momentum fraction $z$ to $z + \Delta z$, $\left<N_q^h(z, z+ \Delta z) \right>$ and expressing it as
 \begin{equation}
\label{EQ_FRAG_MC}
D_q^h(z) \Delta z = \left< N_q^h(z, z+ \Delta z) \right> \equiv  \frac{ \sum_{N_{Sims}} N_q^h(z, z+ \Delta z) } { N_{Sims} }.
\end{equation}

\begin{figure}[phtb]
\centering 
\includegraphics[width=0.8\textwidth]{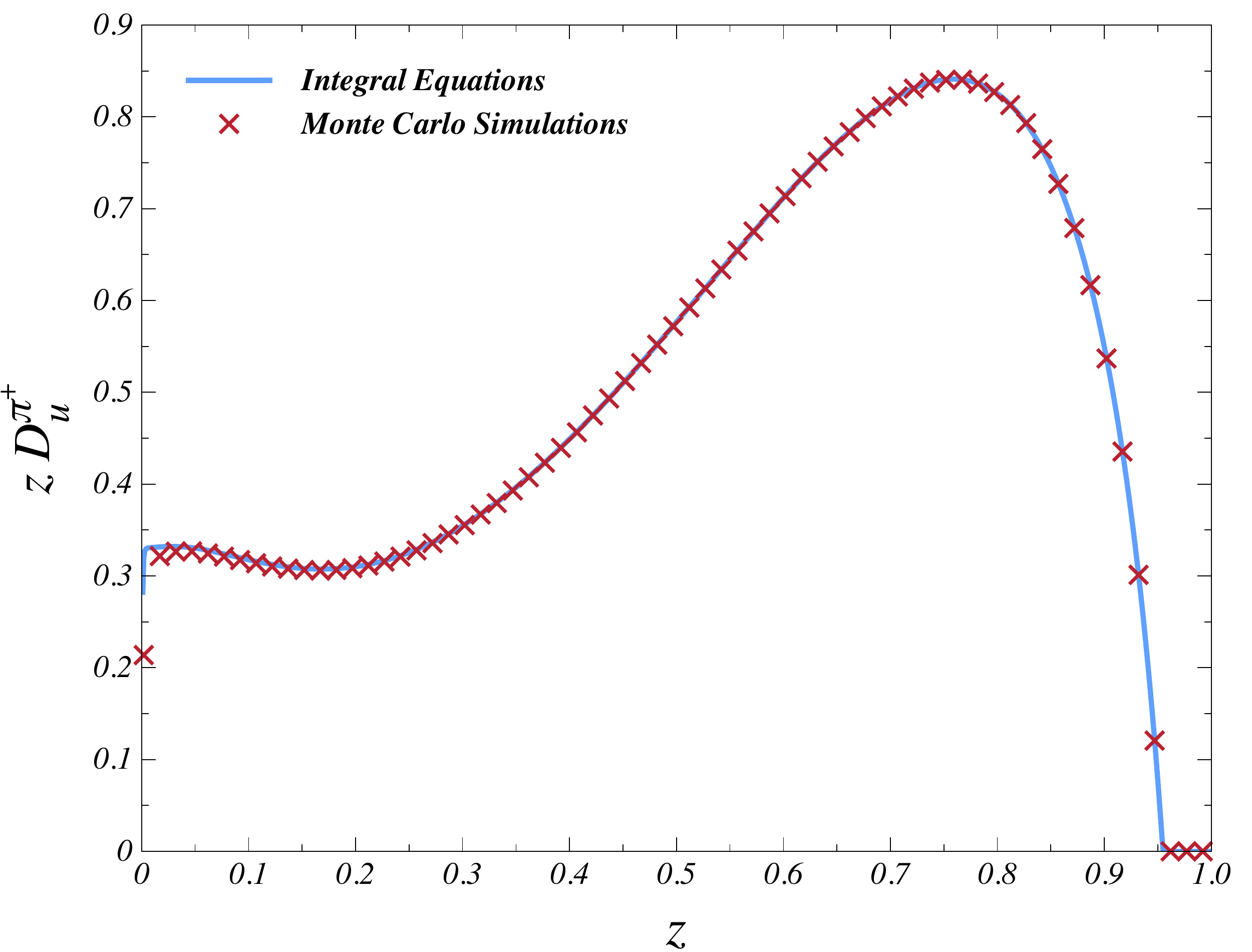}
\caption{Comparison of  the solutions for quark fragmentation function $D_u^{\pi^+}(z) $ in NJL-jet model with only nonstrange pseudoscalar mesons calculated from integral equations, Eq.~(\ref{EQ_FRAG_PROB}) and MC simulation.}
\label{PLOT_MC_INT_COMP}
\end{figure}

 From the construction, it is obvious that the fragmentation functions calculated using the integral equations, Eq.~(\ref{EQ_FRAG_PROB}), should be equivalent to those calculated using the MC method in the limit $N_{Links} \to \infty$ and $N_{Sims} \to \infty$. The plots in Fig.~\ref{PLOT_MC_INT_COMP} show that the solutions for fragmentation functions from both methods are indeed equivalent with a large enough number of emitted hadrons within statistical errors. It follows from Eq.~(\ref{EQ_FRAG_PROB}) that the solution to the integral equations behaves as $zD_q^h(z) \to \mathrm{const}$ as $z \to 0$. This behavior originates in the assumption made by Field and Feynman in the original jet model~\cite{Field:1977fa} that the total fragmentation function can be expressed as an infinite product of elementary fragmentation functions, which allows one to formulate the integral equations. The deviations of the MC solution from the solution of the integral equations for very small $z$ is because in the MC calculation we always use a finite number of hadrons emissions, although in Fig.~\ref{PLOT_MC_INT_COMP} this number was taken to be large enough so as to demonstrate the equivalence to the integral equations.

\begin{figure}[phtb]
\centering 
\includegraphics[width=0.8\textwidth]{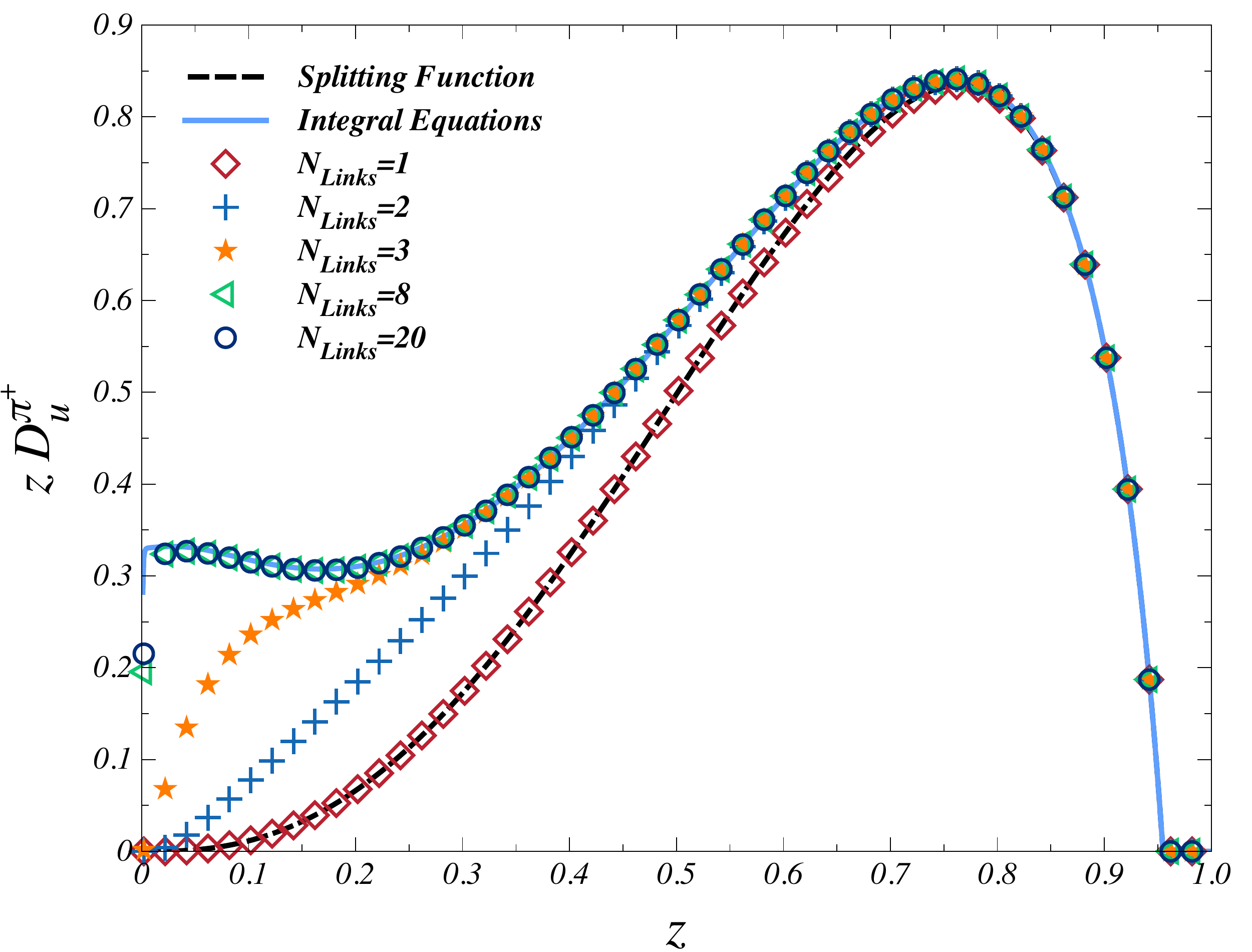}
\caption{The dependence of the solutions for $z D_u^{\pi^+}$ on $N_{Links}$.}
\label{PLOT_FRAG_VS_NLINKS}
\end{figure}

  MC also allows us to study the dependence of the resulting fragmentation functions on the number of hadrons emitted by the quark in the cascade, which could well be relevant to many medium-energy experiments. Figure~\ref{PLOT_FRAG_VS_NLINKS} shows that the solution for $z D_u^{\pi^+}(z) $ with $N_{Links}=1$ [equivalent to the elementary fragmentation function $z \hat{d}_u^{\pi^+}(z)$] is peaked at  $z \sim 0.8$. As we increase the number of emitted hadrons, the solution increases in the low $z$ region because of the hadrons emitted further in the quark jet, where the fragmenting quark typically has a small fraction of the initial quark's light-cone momentum. We can readily see that the solutions saturate after including only a few emitted hadrons, where there is virtually no difference between solutions with $N_{Links} = 8$ and $N_{Links} = 20$, and the discrepancy to the solution of the integral equations only occurs at extremely small values of $z$, approaching the limit $zD(z) \to \mathrm{const}$ in the case of large number of links. Thus we can reliably use the solutions of MC simulations with $N_{Links}\geq 8$.  
  
\subsection{Including the Resonance Decays}  

\begin{figure}[htb]
\centering 
\includegraphics[width=0.6\textwidth]{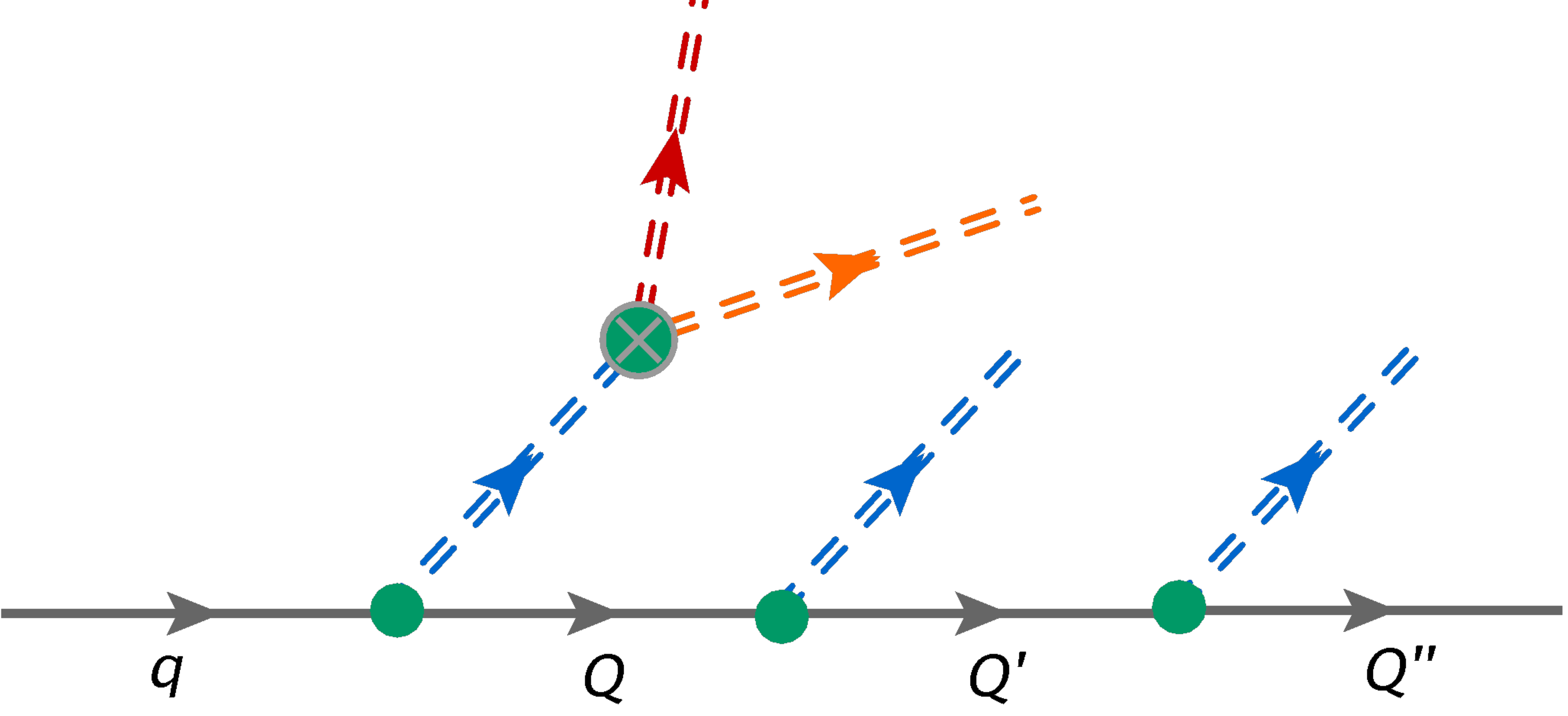}
\caption{Quark-Jet model with the decay of the resonances.}
\label{PLOT_JET_DECAY}
\end{figure}

 In the current version of the NJL-jet model, we included the production of vector mesons, among the other hadrons directly emitted by the quark cascade (so called "primary" hadrons). The vector mesons considered have an extremely short lifetime and decay quickly, thus in an experimental setup one usually detects only their decay products ( "secondary" hadrons), most often pions and kaons. Schematically, the process is depicted in Fig.~\ref{PLOT_JET_DECAY}. Consequently, to best describe the experimentally measured fragmentation functions of a hadron $h$ from the Eq.~(\ref{EQ_FRAG_MC}), we should not only consider the number of primary hadrons with a given range of  light-cone momentum fraction of the original quark, $N_q^h(z, z+ \Delta z)$, but also add the number of hadrons $h$ satisfying the above criteria that originate from the decays of primary vector mesons (or in general other hadronic resonances). This is accomplished using the dependence of the decay probability of a hadronic resonance $h$ on the fractions of its light-cone momentum,  $z_1$ to $z_1+d z_1,..., z_n$ to $z_n+dz_n; \sum_i z_i =1$,  carried by the decay products  $h_1, ... h_n$, denoted as  $dP^{h\rightarrow h_1... h_n}(z_1,...,z_n)$. First we perform the MC simulations described in the previous section and record all the produced (primary) hadrons along with the fractions of the initial quark's light-cone momentum. We calculate the fragmentation functions without the decays using the formula in Eq.~(\ref{EQ_FRAG_MC}). Second, we consider each encountered resonance particle $h$ with a  momentum fraction of the initial quark $z$ and its possible strong decay channels, randomly selecting one according to the corresponding relative branching ratio. Then we randomly generate the light-cone momentum fractions $z_1,..., z_n$ carried by the decay products $h_1, ... h_n$ according to the probability $dP^{h\rightarrow h_1... h_n}(z_1,...,z_n)$. The decaying hadron $h$ is removed from the list of the produced hadrons and the decay products $h_1,..., h_n$ are added with their respective fractions of the initial quark's light-cone momenta $z z_1, ..., z z_n$.  The fragmentation functions are once again calculated using the new list of produced hadrons using the formula in Eq.~(\ref{EQ_FRAG_MC}). (In general, the decay of a primary resonance can produce another resonance of a lower mass that decays itself, so we start the simulation of the decay process from the heaviest resonances present and move on to the lighter ones.)
 
 In the current article, we consider only the strong decays of the vector mesons to two-body pseudoscalar meson final states  for simplicity. A generalization to  include the three or more-body final states is straightforward, although it requires sampling the nontrivial phase space factors using Monte Carlo techniques. We consider all of the two-body strong decays of $\phi$, $K^*$, and $\rho$ mesons with the corresponding relative branching ratios as given in the "Review of Particle Physics"~\cite{Nakamura:2010zzi}. For calculation of the two-body decay probability function, let us consider the initial hadron $h$ with mass $m_h$, momentum $q$ decaying to hadron $h_1$ with a mass $m_{h1}$, and a momentum $p_1$ and hadron $h_2$ with mass $m_{h_2}$, and momentum $p_2$.  We also denote the light-cone momentum fraction of the $h$ carried by $h_1$ as $z_1 \equiv p_{1}^-/q^- $ and the fraction carried by $h_2$ as $z_2 \equiv p_2^-/q^-$, where trivially $z_1+z_2=1$. Thus, the decay probability is  a function of only one momentum fraction chosen to be the $z_1$. The $dP^{h\rightarrow h_1,h_2}(z_1)$ is determined as a product of the decay amplitude squared times a two-body phase space factor.
  
 A detailed description of the decay amplitudes and the branching ratios into different channels can be calculated using specific models ( for example model Lagrangians for the interaction from Ref. \cite{Matsuyama:2006rp}). Here, we are only concerned with the dependence of the decay probability on $z_1$, where we average over the polarization of the vector meson. Thus, the Lorentz invariance requires that amplitude squares depend only on scalar products of the 4-momenta of the particles involved in the decay, which in turn are trivially  expressed through on-mass-shell conditions in terms of their masses. Thus, the only dependence on $z_1$ comes from the two-body phase space factor. Our goal is to express the elementary probability for the decay as a function of $z_1$, assuming a constant $C_h^{h_1 h_2}$ for the amplitude squared of the decay. For that we integrate over all components of momenta $p_1$ and $p_2$ with exception of $p_1^-$, assuming $\bf{q_\perp=0}$ and using the light-cone form for the two-body phase space factor:  
 \begin{align}
\label{EQ_DIFF_CS_LF}
dP^{h\rightarrow h_1,h_2}(z_1) =& C_h^{h_1 h_2}\  d p_1^- \int \frac{ d^2 \mathbf{p}_{1\perp}}{(2\pi)^3 2 p_1^-} \frac{d p_2^-  d^2 \mathbf{p}_{2\perp}}{(2\pi)^3 2 p_2^-} (2\pi)^4 \delta^4(q-p_1-p_2)\\
=& \frac{C_h^{h_1 h_2}}{8 \pi}\  dz_1 \int_0^\infty d l\  \delta \left(l - \left[z_1 z_2\ {m_h^2} - z_2 m_{h1}^2 - z_1 m_{h2}^2\right] \right)|_{z_2=1 - z_1}\\
= &\left\{ 
  \begin{array}{l l}
   \frac{C_h^{h_1 h_2}}{8\pi}\ d z_1  & \quad \text{ if $z_1 z_2\ {m_h^2} - z_2 m_{h1}^2 - z_1 m_{h2}^2 \geq 0; \ z_1+z_2=1$,}\\
    0 & \quad \text{otherwise.}\\
  \end{array} \right.
\end{align}

Here we note that the numerical values for the $z_1$ ranges for various decays exactly match those shown in the plots in Fig. 2 of Ref.~\cite{Andersson:1977xs}. We determine the $C_h^{h_1 h_2}$ such that the total probabilities  $\int dP^{h\to h_1 h_2}$ of  the decays to different pairs $\{h_1 h_2\}$ relate as the corresponding relative branching ratios and the normalization condition $\sum_{\{h_1 h_2\}}\int dP^{h\to h_1 h_2}=1$ is satisfied. 
 
 \section{Results}
\label{SEC_RES}
  
\begin{figure}[htb]
\centering 
\includegraphics[width=0.8\textwidth]{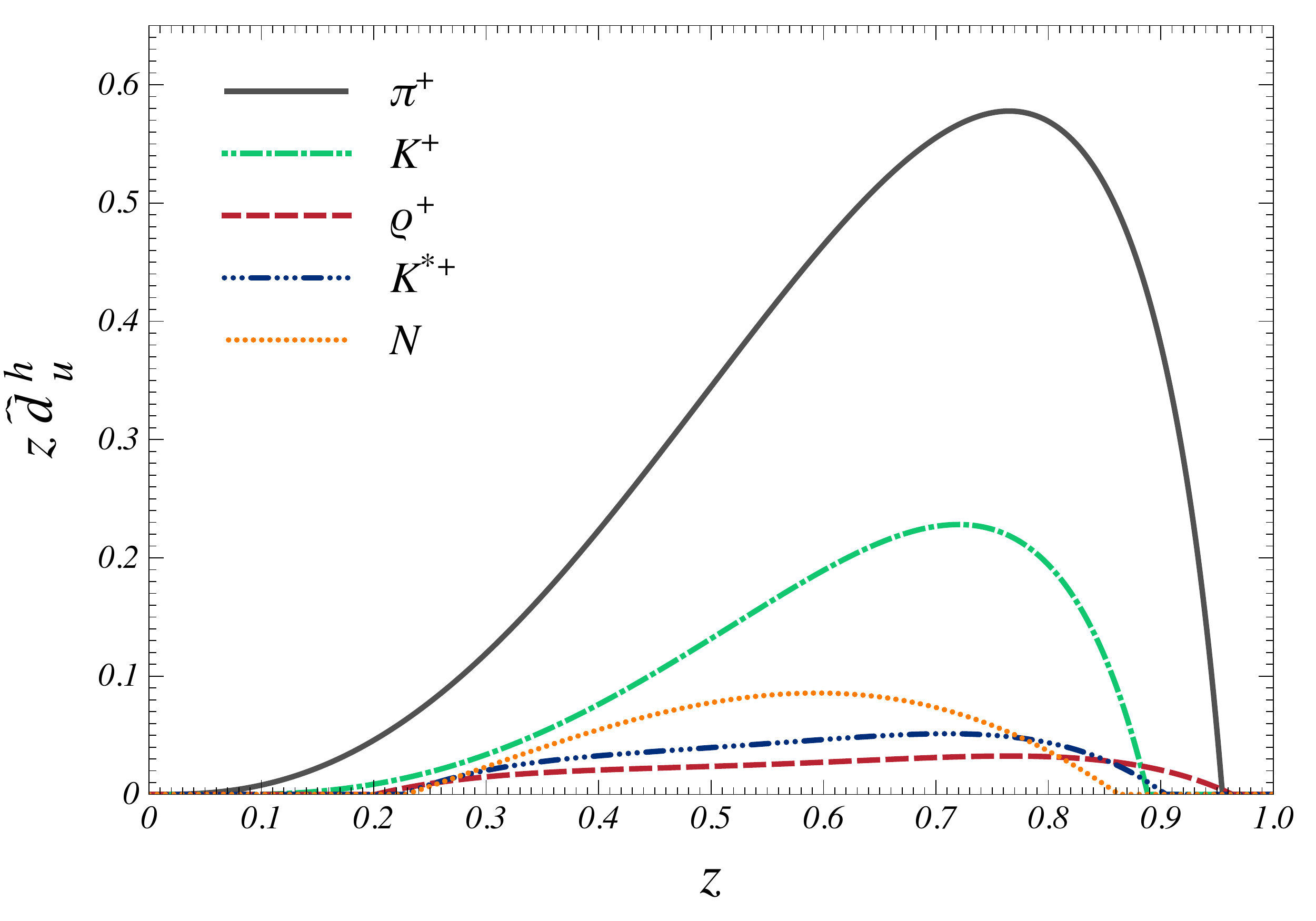}
\caption{Elementary fragmentation functions for $u$ quark, $z \hat{d}_u^{h}$,  for all included emission channels.}
\label{PLOT_FRAG_DRV_NNB}
\end{figure}

   The plots in Fig.~\ref{PLOT_FRAG_DRV_NNB} depict the elementary fragmentation functions, $z \hat{d}_u^h(z)$ from a $u$ quark for pseudoscalar,  vector meson and nucleon emission with the mass of scalar antidiquark $M_{D}=0.687~\mathrm{GeV}$ from Ref.~\cite{Cloet:2007em} and the following quark-hadron couplings (calculated in Appendix \ref{APP_QM_COUPLING}): $g_{\pi qQ}=3.15,\ g_{K qQ}=3.387,\ g_{\rho qQ}=1.5$, $g_{K^{*} qQ}=1.91$, and $g_{\phi qQ}=2.29$. 
 
 In order to compare our results with experimental measurements  or empirical parametrizations we evolve them at next-to-leading order (NLO) from our model scale $Q_{0}^{2}=0.2~{\rm GeV}^{2}$ using the software from Ref.~\cite{Botje:2010ay}. The details of determination of the model scale are given in \cite{Matevosyan:2010hh}.

\begin{figure}[phtb]
\centering 
\subfigure[] {
\includegraphics[width=0.48\textwidth]{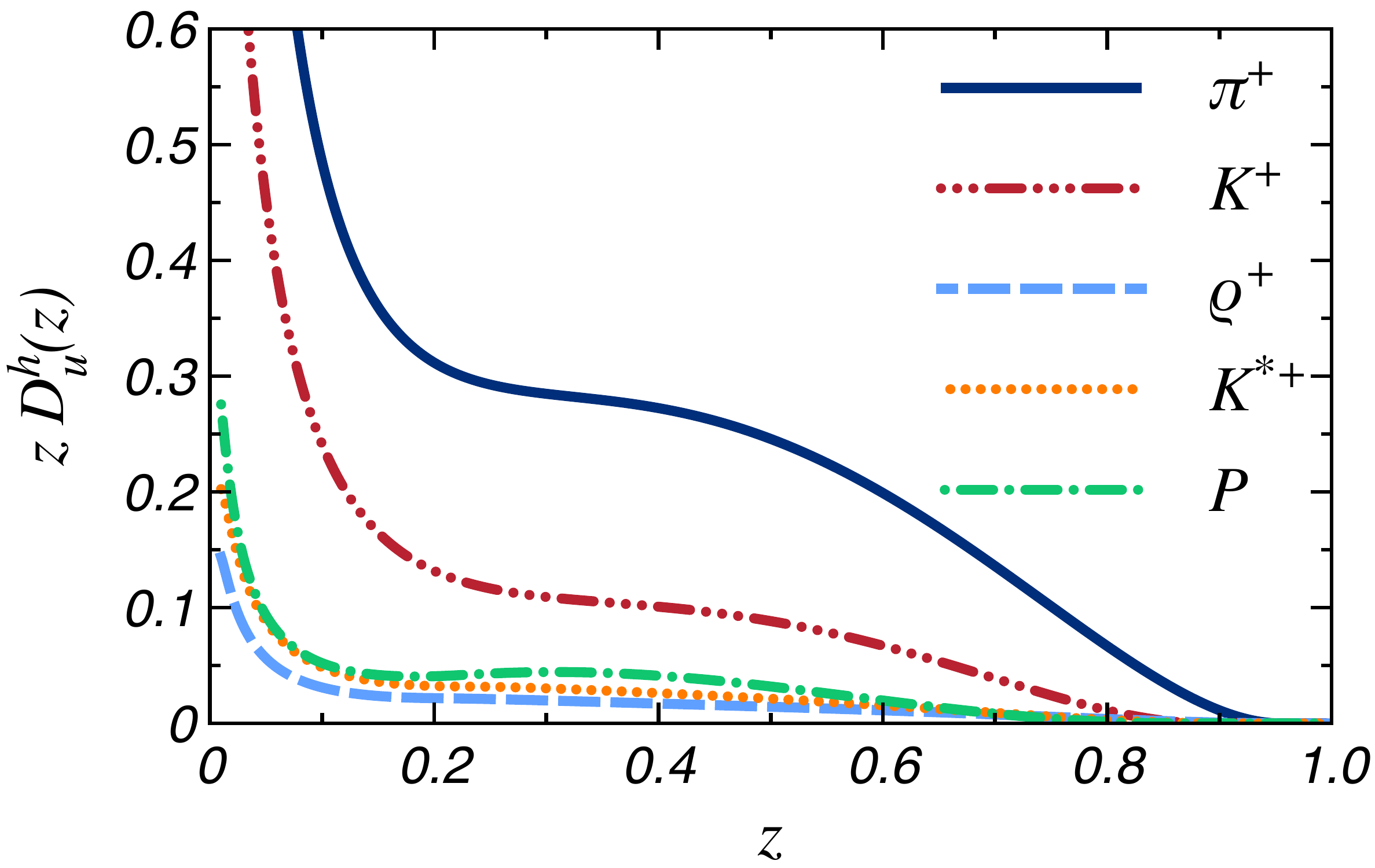}}
\hspace{0.1cm} 
\subfigure[] {
\includegraphics[width=0.48\textwidth]{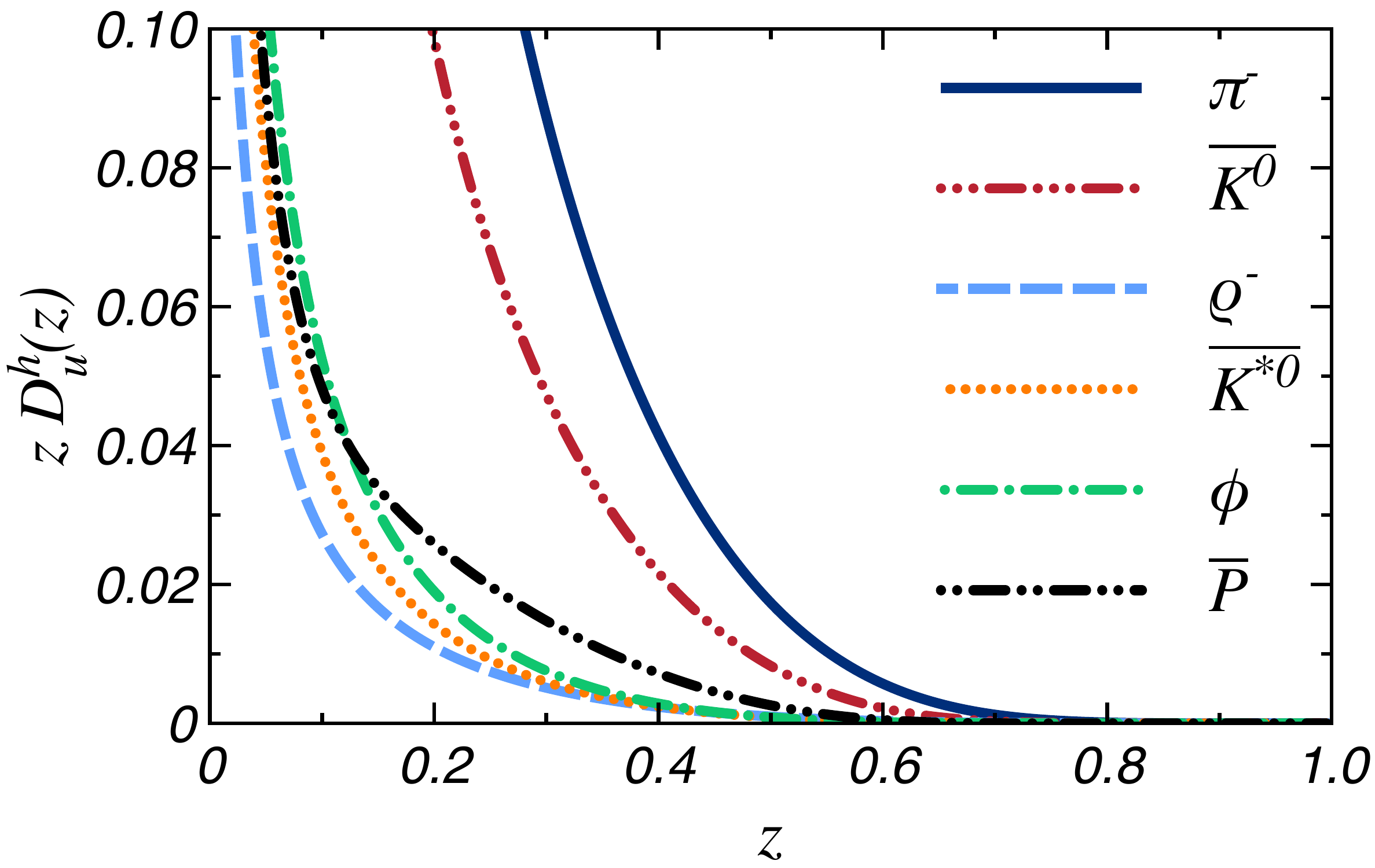}
}
\caption{The solutions for  (a) favored and (b) unfavored fragmentation functions from $u$ quark to various hadrons  evolved at NLO to $Q^2=4~\rm{GeV^2}$. }
\label{PLOT_FRAG_HADRONS}
\end{figure}

The solutions for the favored and unfavored fragmentation functions from an $u$ quark to primary hadrons (without any resonance decays), evolved at NLO to a typical scale $Q^2=4~\rm{GeV}^2$, are shown in the plots in Fig.~\ref{PLOT_FRAG_HADRONS}. Here, we note that the fragmentation to nucleons is comparable to the case of vector mesons.  
 
\begin{figure}[phtb]
\centering 
\subfigure[] {
\includegraphics[width=0.9\textwidth]{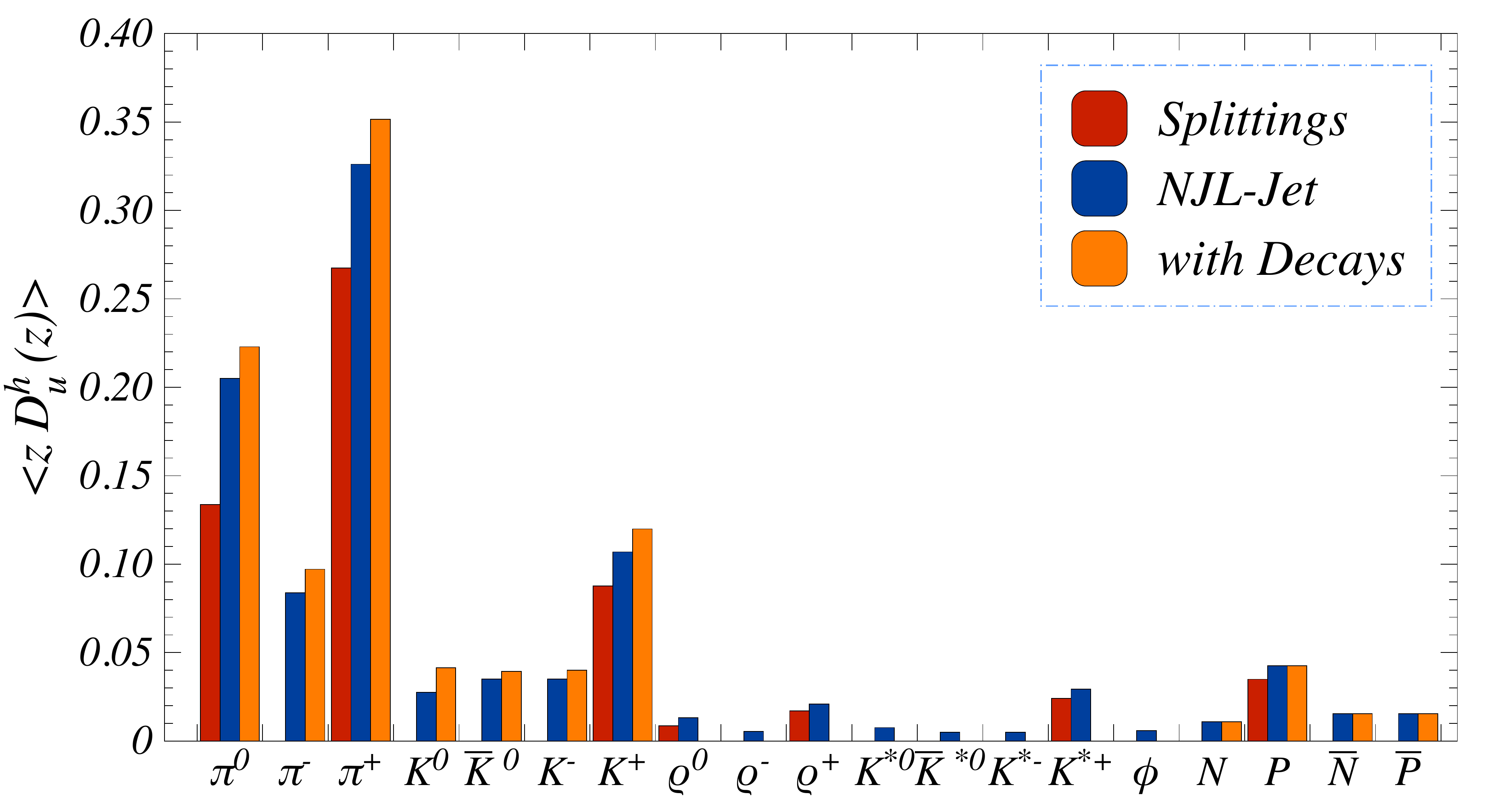}}
\vspace{0.1cm} 
\subfigure[] {
\includegraphics[width=0.9\textwidth]{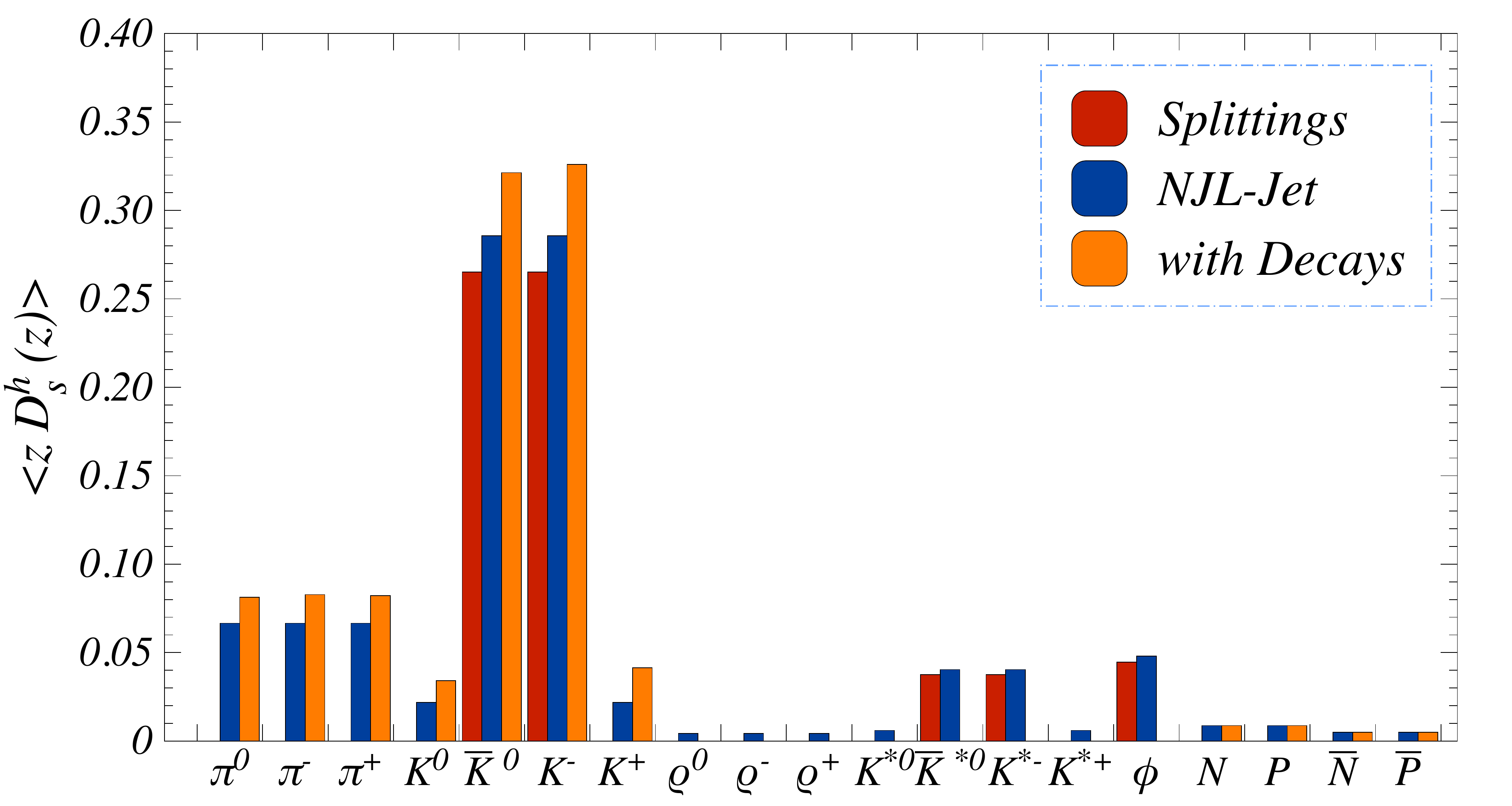}
}
\caption{The total fractions of momenta carried by mesons of type $m$ from the jet of  (a) $u$ and (b) $s$ quarks calculated using the numerical solutions for fragmentation functions at the model scale $Q_0^2=0.2~\rm{GeV^2}$. Here "splittings" denote the momentum fractions calculated using the elementary splitting functions $\langle z \hat{d}_q^m(z) \rangle$,  "NJL-jet" and "with decays" denote the momentum fractions calculated from solutions for the fragmentation functions without and with inclusion of pions and kaons originating from decays of vector meson resonances (the corresponding columns arranged from left to right for each hadron). }
\label{PLOT_MOM_SUM}
\end{figure}

 The plots in Fig.~\ref{PLOT_MOM_SUM} depict the total light-cone momentum fractions of the initial quark carried by the emitted hadrons of different species, calculated at model scale of $Q_0^2=0.2~\mathrm{GeV}^2$. (As discussed in Ref.~\cite{Matevosyan:2010hh}, the momentum and isospin sum rules can only be reliably satisfied at model scale, as the NLO QCD evolution kernels have known singularities in the low $z$ region, in practice leaving the values of the fragmentations for $z \lesssim 0.01$ undetermined.)   Here we compare the fractions for elementary fragmentation functions, the solutions without the resonance decays and the solutions after the resonance decays. The sum of the fractions from elementary splitting functions calculated in the NJL model amounts to about half of the initial quark's light-cone momentum, where the rest is carried by the remnant quark. In the quark-jet picture,  we sum up the light-cone momentum fractions of all hadrons emitted in the chain. This implies that the initial quark transfers all of its light-cone momentum to the produced hadrons, which we confirm in our numerical solutions (the sum of all the fractions is 1 within numerical errors). This shows that after resonance decays, the pions carry approximately $67\%$ and kaons carry about $24\%$ of the initial $u$ (or $d$) quark's  light-cone momentum. In the case of an initial $s$-quark, pions carry approximately $25\%$ and kaons carry about $72\%$ of its light-cone momentum.  The rest of the momentum is carried by nucleons and antinucleons, where the momentum sum rule is satisfied within $0.1\%$.

\begin{figure}[htb]
 \centering 
\subfigure[] {
\includegraphics[width=0.48\textwidth]{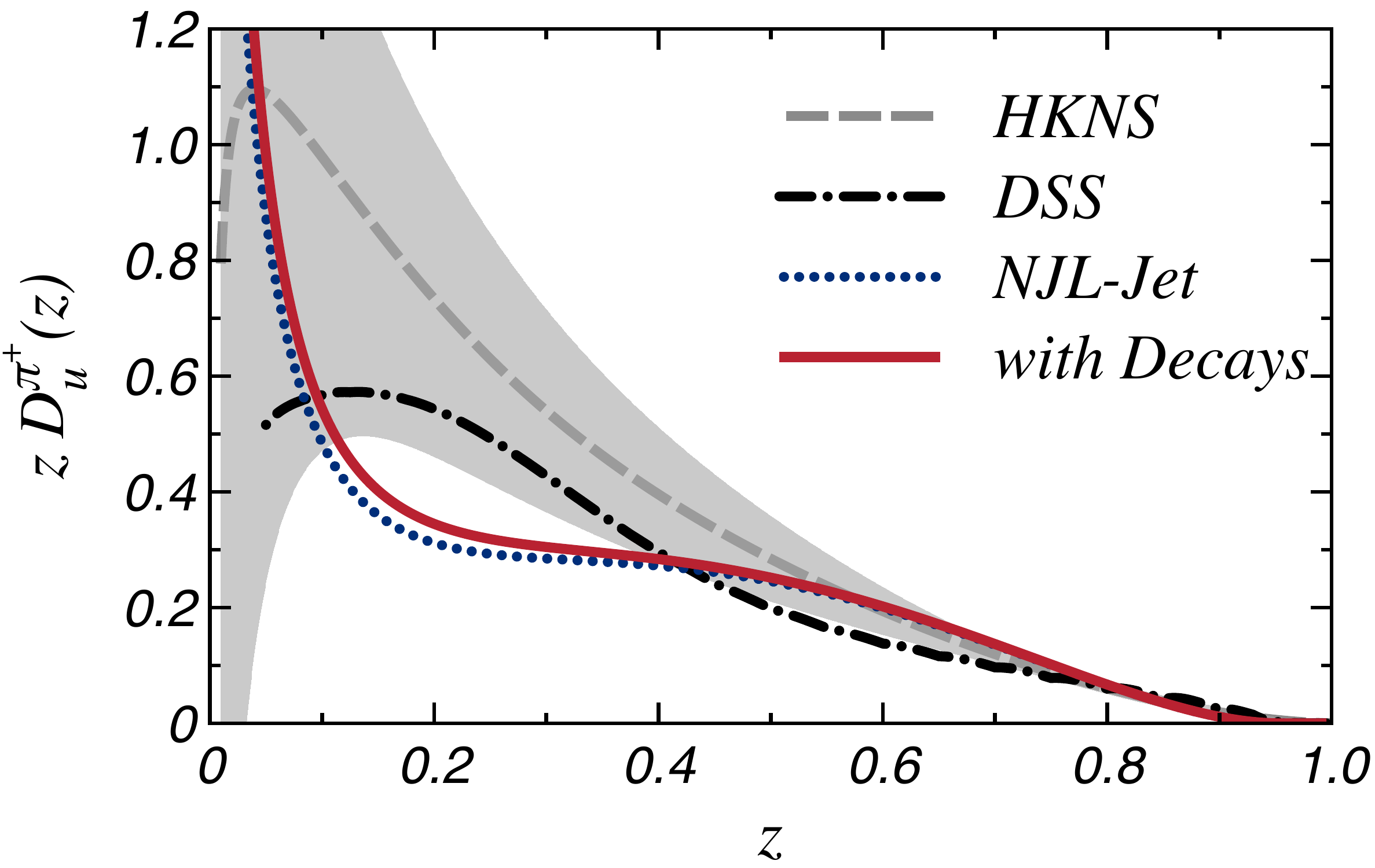}}
\hspace{0.1cm} 
\subfigure[] {
\includegraphics[width=0.48\textwidth]{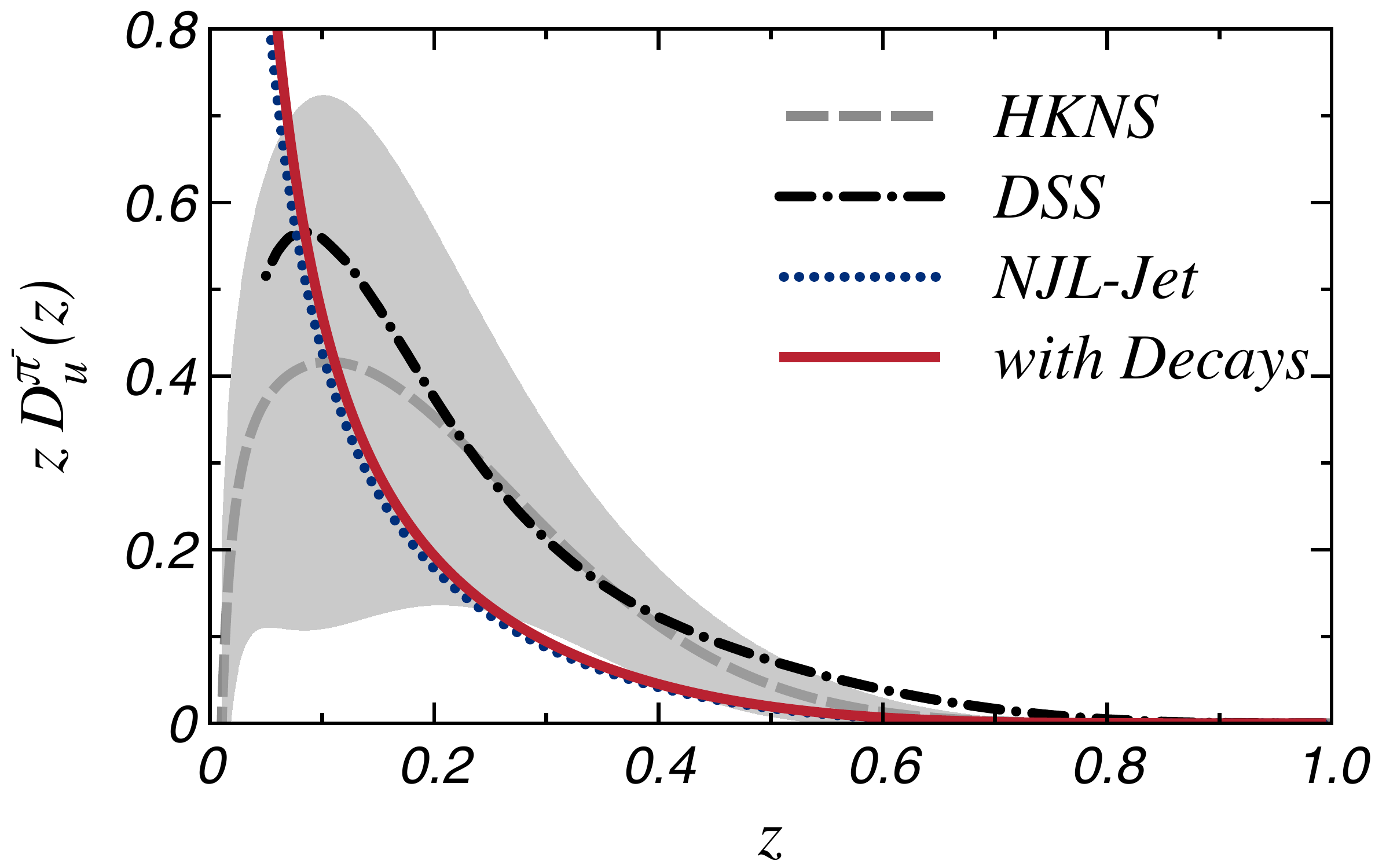}
}
\caption{The solutions for fragmentation function $z D_u^{\pi}(z)$ evolved at NLO to $Q^2=4~\rm{GeV^2}$. The results are compared to the empirical parametrizations of the experimental data from Refs.~\cite{Hirai:2007cx} (HKNS) and~\cite{deFlorian:2007aj} (DSS). Here "NJL-jet" and "with decays" denote the fragmentation functions calculated without and with inclusion of pions originating from decays of vector meson resonances. The shadows show the uncertainties of the empirical functions of Ref.~\cite{Hirai:2007cx}.}
\label{PLOT_FRAG_UPI_DECAY}
\end{figure}

\begin{figure}[htb]
 \centering 
\subfigure[] {
\includegraphics[width=0.48\textwidth]{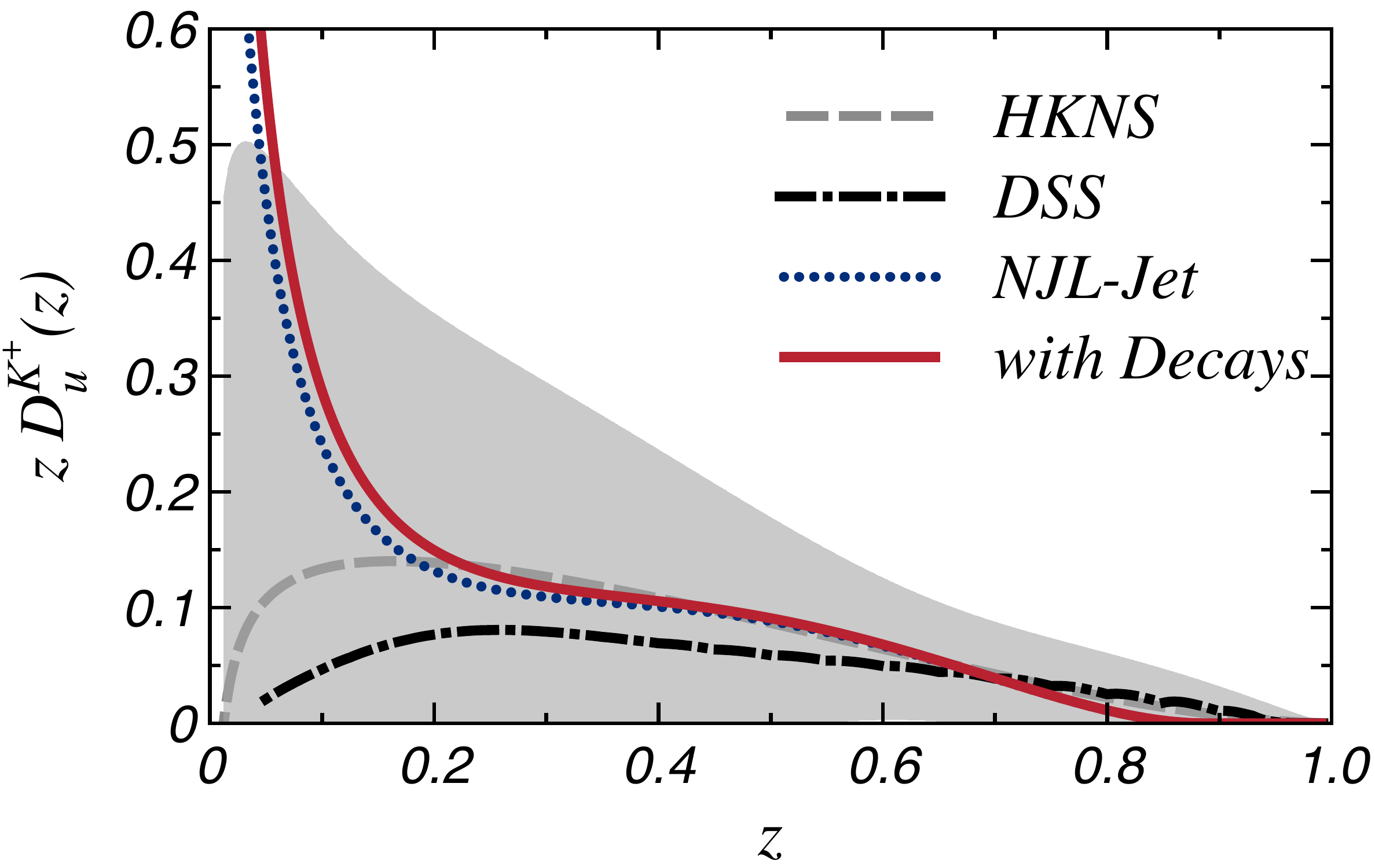}}
\hspace{0.1cm} 
\subfigure[] {
\includegraphics[width=0.48\textwidth]{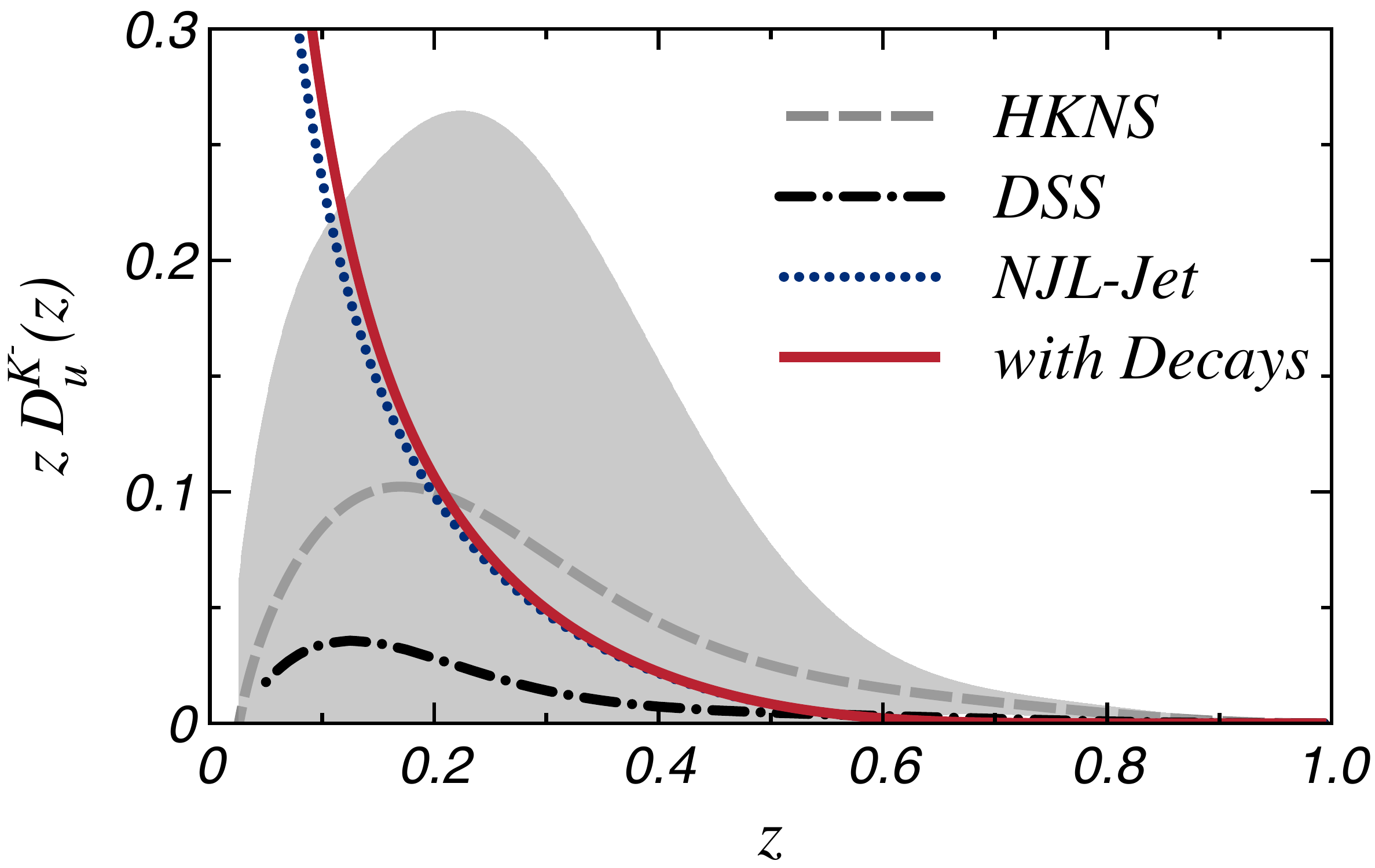}
}
\caption{Same as Fig.~\ref{PLOT_FRAG_UPI_DECAY} for the case $z D_u^{K}(z)$.}
\label{PLOT_FRAG_UK_DECAY}
\end{figure}

\begin{figure}[htb]
 \centering 
\subfigure[] {
\includegraphics[width=0.48\textwidth]{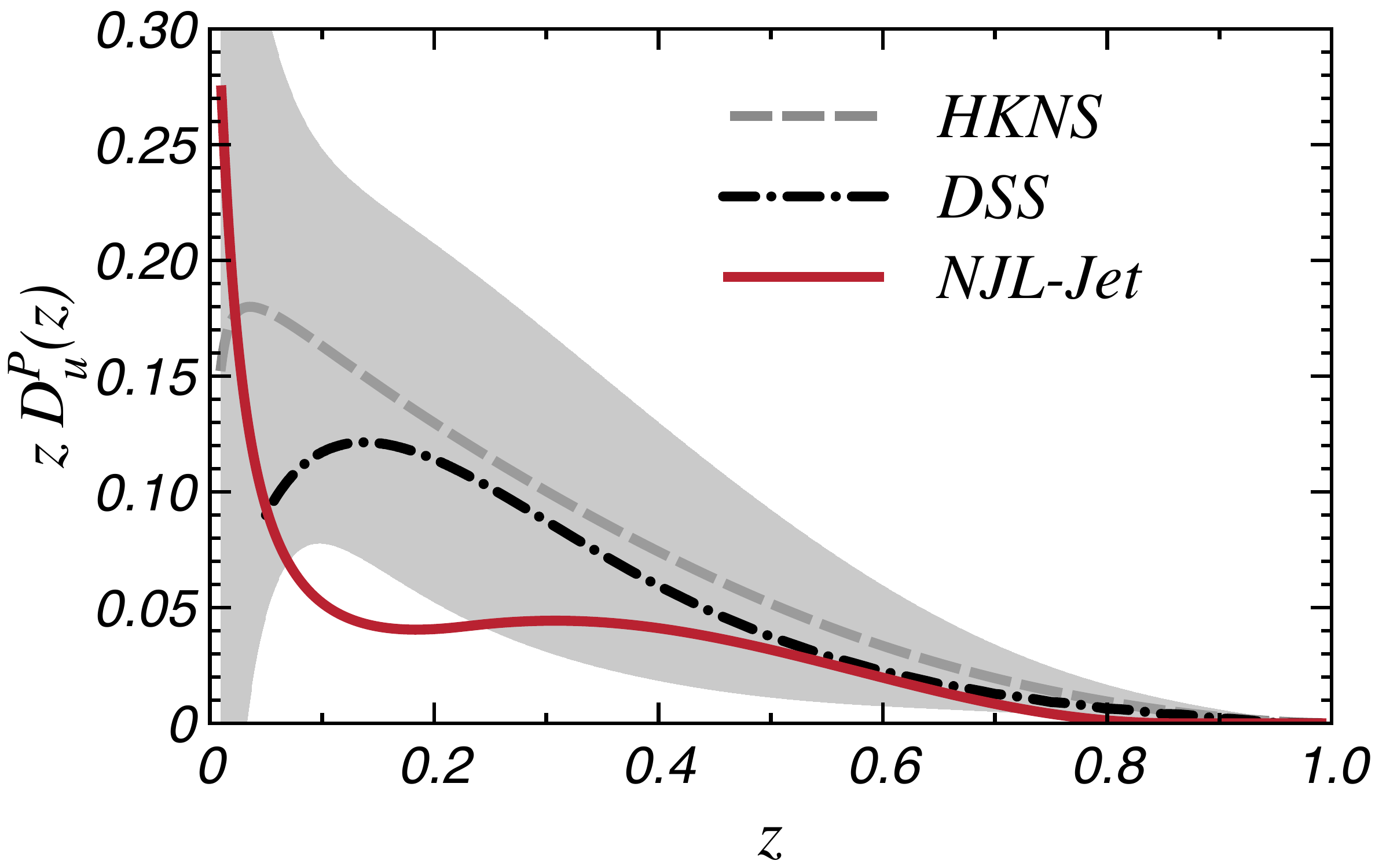}}
\hspace{0.1cm} 
\subfigure[] {
\includegraphics[width=0.48\textwidth]{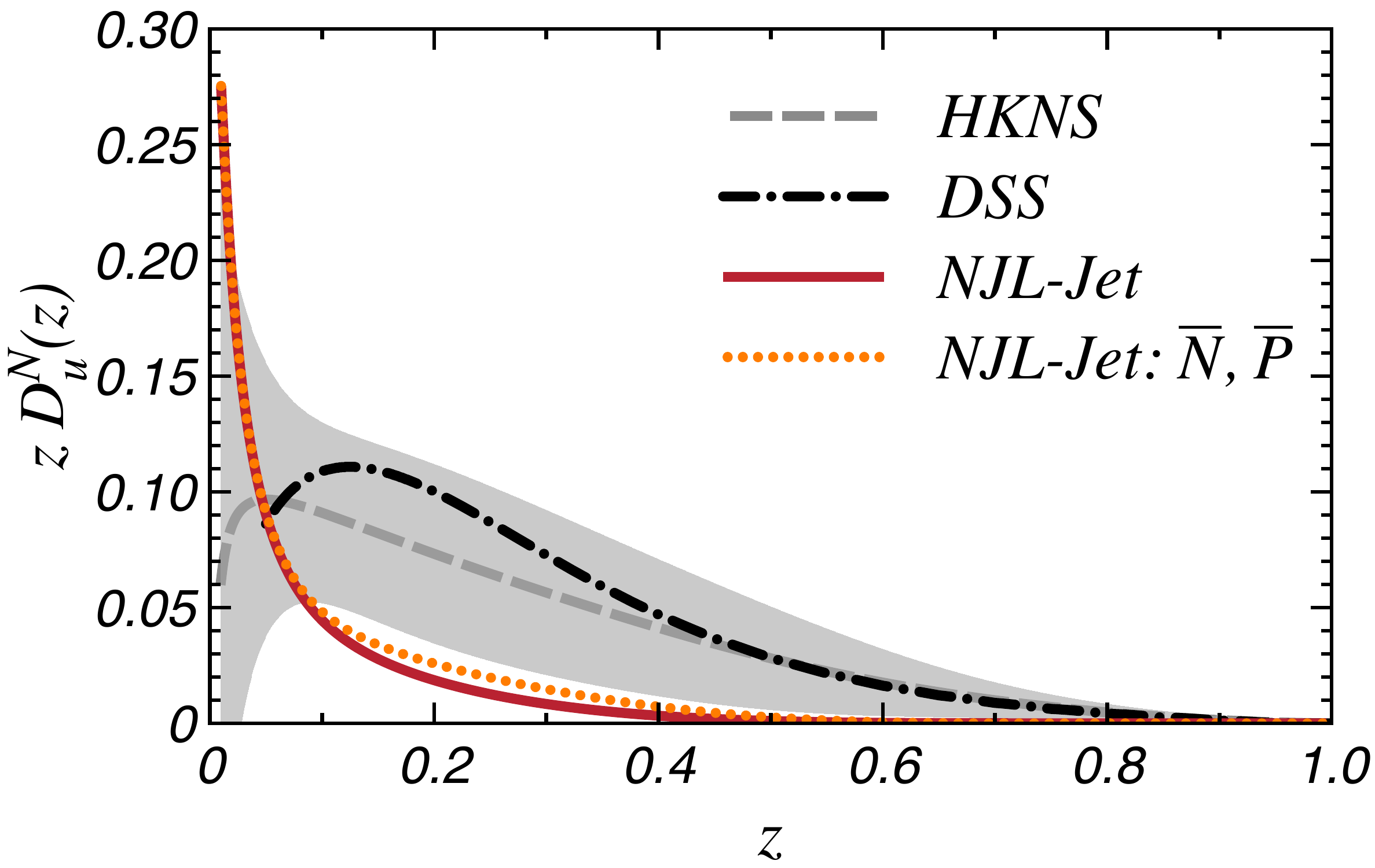}
}
\caption{Same as Fig.~\ref{PLOT_FRAG_UPI_DECAY} for the case $z D_u^{P}(z)$ and $z D_u^{N}(z)$.}
\label{PLOT_FRAG_UPN_DECAY}
\end{figure}

The results for the solutions for fragmentation function $z D_u^{\pi^+}$ and $z D_u^{\pi^-}$ prior to and after the vector meson decays are shown in Fig.~\ref{PLOT_FRAG_UPI_DECAY}. Here, we see that the inclusion the resonance decay slightly changes the shape of the function in mid- to low-$z$ region, bringing it closer to the empirical parametrizations of the experimental data from Refs.~\cite{Hirai:2007cx} (HKNS) and~\cite{deFlorian:2007aj} (DSS). Similarly, the results for  $z D_u^{K}$  prior to and after the vector meson decays are shown in Fig.~\ref{PLOT_FRAG_UK_DECAY} and the results for $z D_u^{P}$ and $z D_u^{N}$ are shown in Fig.~\ref{PLOT_FRAG_UPN_DECAY}. For $z D_u^{P}$ the agreement with the empirical parametrizations from Refs.~\cite{Hirai:2007cx} (HKNS) and~\cite{deFlorian:2007hc} (DSS) is reasonable, while our results for $z D_u^{N}$ are well below the empirical curves. This is because we only include the scalar diquarks in our model, making the fragmentation of a $u$ quark to $N$ an unfavored process, while in the empirical parametrizations it is usually assumed $D_u^P(z)\simeq 2 D_u^N(z)$ based on naive $SU(2)$ flavor symmetry arguments. In the future developments of the NJL-jet model, the inclusion of the axial-vector diquark intermediate states in the quark to nucleon splitting process will include a favored channel [depicted in Fig.~\ref{PLOT_FRAG_Q_N}a ] for emission of a neutron from a $u$ quark.

\begin{figure}[htb]
 \centering 
\subfigure[] {
\includegraphics[width=0.48\textwidth]{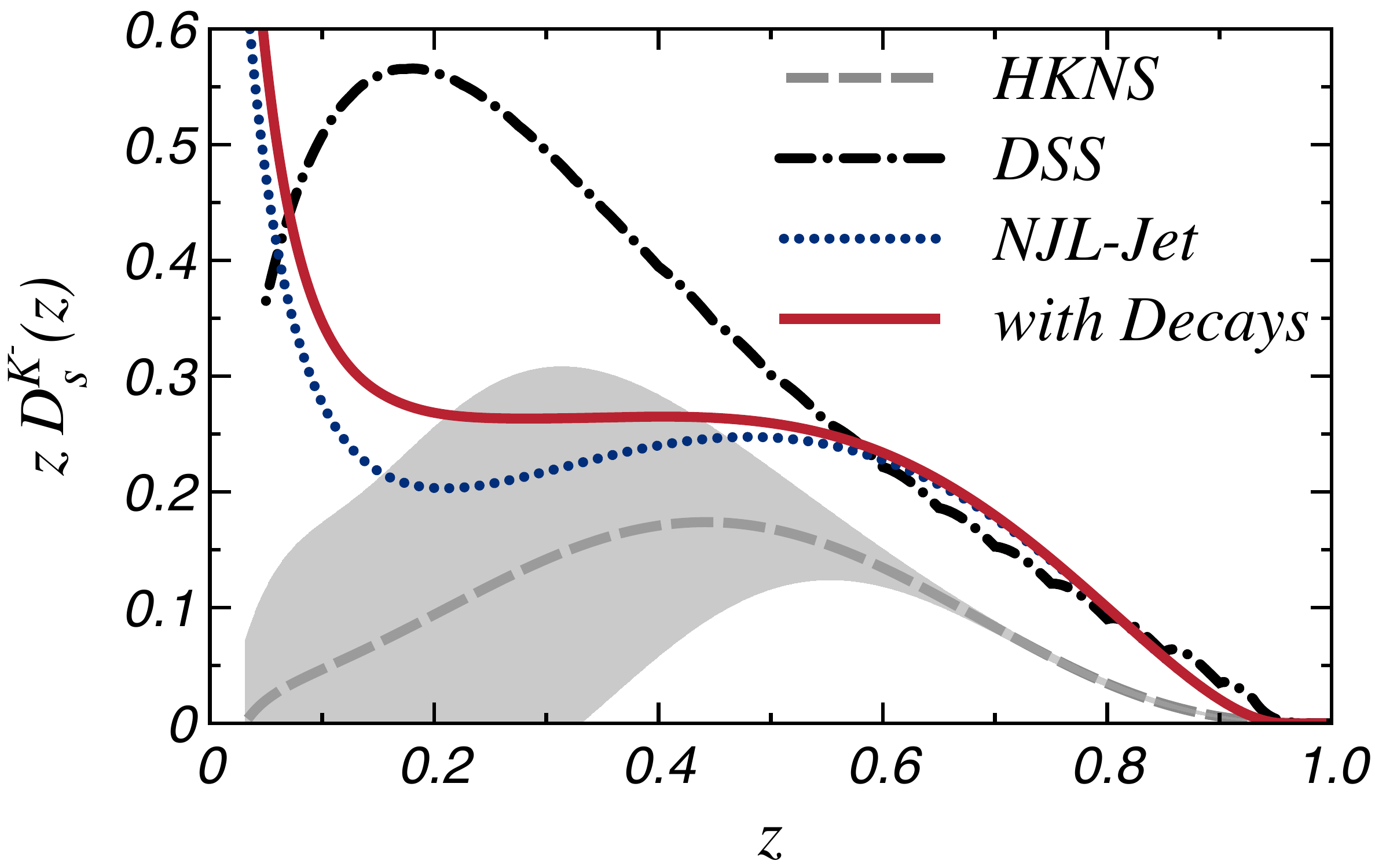}}
\hspace{0.1cm} 
\subfigure[] {
\includegraphics[width=0.48\textwidth]{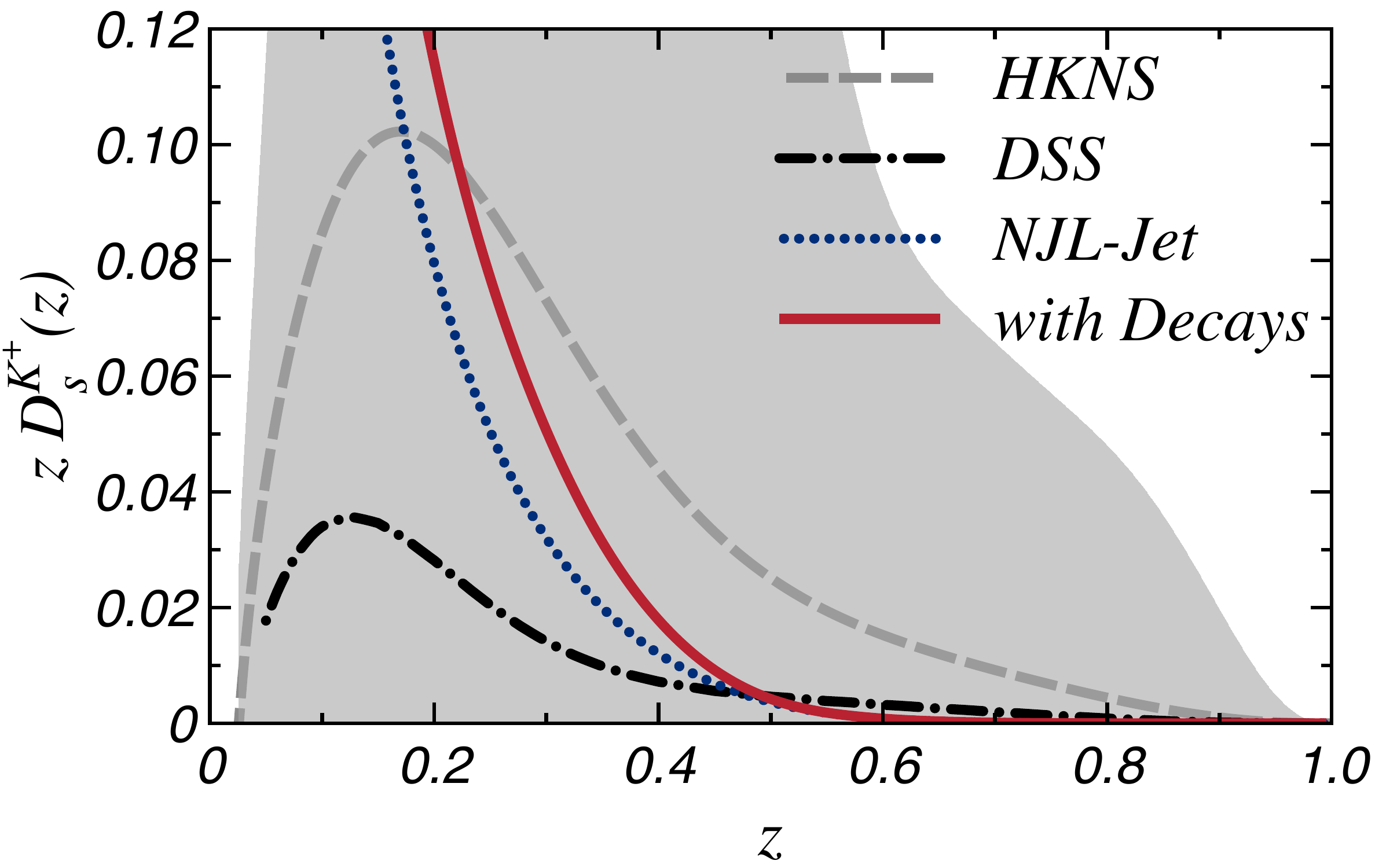}
}
\caption{Same as Fig.~\ref{PLOT_FRAG_UPI_DECAY} for the case $z D_s^{K}(z)$.}
\label{PLOT_FRAG_SK_DECAY}
\end{figure}

The solutions for the $z D_s^{K^-}$ and $z D_s^{K^+}$ are shown in Fig.~\ref{PLOT_FRAG_SK_DECAY}. Here we see the strong discrepancies in the global fits \cite{Hirai:2007cx} and \cite{deFlorian:2007aj} of the experimental data that illustrates the need for the model calculations providing additional insight into the quark fragmentation process.

\section{Conclusions and Outlook}
\label{SEC_CONC}

 In the current article, we added the vector meson, nucleon, and antinucleon emission channels to NJL-jet framework for calculating quark fragmentation functions. We also included the two-body decays of the vector mesons to pseudoscalars, which allows for easier comparison of the calculated fragmentation functions with the experimental measurements or their parametrizations. We employed the Monte Carlo method to obtain the fragmentation functions in a quark-cascade description. Here we demonstrated that the Monte Carlo approach to calculating the fragmentation functions in NJL-jet framework is a powerful and reliable method. We reproduced the fragmentation functions calculated as solutions of the previously employed integral equations, where only the light quarks and pions were included. Moreover, we showed that the MC approach allows for the flexibility to surpass the model limitations necessary in formulating the integral equations. That is, in the future MC studies we can assume an initial quark carrying only a finite momentum and thus emitting a finite number of hadrons. We demonstrated that the medium and low $z$ regions of the fragmentation functions are greatly affected by the number of the emitted hadrons, thus the finiteness of the quark momentum might have a noticeable effect. The future development of the NJL-jet model would also allow an access to the transverse momentum distribution of the produced hadrons, thus becoming relevant for the analysis of a large variety of semi-inclusive data.  The MC approach provides a robust and efficient platform for implementing these and other possible extensions of the NJL-jet model that would allow for a much more detailed description of the physical picture.  
 
  A further advantage of the MC approach is in reducing the numerical task in solving for the fragmentation functions when including many more channels for emitted hadrons. Here, solving the integral equations requires inverting larger and larger matrices, while the MC procedure can be drastically sped up by trivially parallelizing the task and  solving simultaneously on computer clusters.

 The results for the fragmentation functions exhibit only slight changes with addition of the new hadronic channels. In particular, vector meson-quark couplings are relatively small in our model, lowering their relative contribution to the fragmentation process. The resulting quark fragmentation functions exhibit a good agreement with the empirical parametrizations in high $z$ region, staying reasonably close to them also in mid and low $z$ regions in part due to the slight modifications of the shape of the functions in this region brought by the effects of the vector meson decays. On the other hand, the plots in  Fig.~\ref{PLOT_MOM_SUM} show that  pions, kaons, nucleons, and antinucleons include the most dominant contributions in fragmentation process, thus the inclusion of higher resonance states is unlikely to be important. A notable underestimation of the fragmentation of $u$ quark to neutrons, shown on plots in Fig.~\ref{PLOT_FRAG_UPN_DECAY}b, demonstrates the need to include the axial-vector diquark for a more realistic description of the fragmentations to nucleons, which will be accomplished in the future developments of the model.
 
  In this work, we followed our previous NJL-jet calculations and used the LB scheme to calculate the regularized fragmentation functions. However, as we have pointed out, the inclusion of vector mesons in the NJL model favors the PT regularization scheme, which is free of unphysical decay thresholds. Here, we have used this scheme only to assess the vector meson-quark couplings, but in future work on extensions of the model we will use the PT scheme consistently throughout.
  
 The  future development of the NJL-jet model using the Monte Carlo framework will allow us to study the  transverse momentum dependence of the quark fragmentation functions as well as polarized fragmentation functions.  This can be accomplished within the NJL framework, without inclusion of any additional parameters.
 
  Last, the medium modifications of the fragmentation functions may be essential to our understanding of the semi-inclusive processes on nuclear targets. Medium effects have long been studied in the NJL model, yielding a successful description of modifications of nucleon properties in nuclei~\cite{Bentz:2001vc,Mineo:2003vc, Cloet:2005pp,Cloet:2006bq}.  Thus the model should provide a reliable framework for the calculation of medium modifications of quark hadronization process.

 \section*{Acknowledgements}
 
This work was supported by the Australian Research Council through Grant No. FL0992247 (A.W.T.), by the University of Adelaide and by a Subsidy for Activating Educational Institutions from the Department of Physics, Tokai University.

\appendix

\section{Flavor Factors for Distribution and Splitting Functions}

The flavor factors given in Table~\ref{TB_FLAVOR_FACTORS} are calculated from the corresponding $SU(3)$ flavor matrices (for details see, e.g., Ref.~\cite{Klimt:1989pm}). With the interaction Lagrangian in the vector channel given in Section~\ref{SEC_VM}, the $\phi$ meson is a pure $s\bar{s}$ state, i.e. an ideal mixture of flavor singlet ($\lambda_0$) and octet ($\lambda_8$).

\begin{table}[hpt]
\caption{Flavor Factors $C_q^m$}
\label{TB_FLAVOR_FACTORS}
\begin{center}
\resizebox{0.9\textwidth}{!}{
\begin{tabular}{|c | c|c|c|c|c|c|c|c|c|c|c|c|c|c|c|} 
\hline  $C_q^m$ & $\pi^0$ & $\pi^+$ & $\pi^-$ & $K^0$ & $\overline{K}^0$ & $K^+$ & $K^-$ & $\rho^0$ & $\rho^+$ & $\rho^-$ & $K^{*0}$ & $\overline{K}^{*0}$ & $K^{*+}$ & $K^{*-}$ & $\ \phi\ $  \\
\hline $u$ & 1 & 2 & 0 & 0 & 0 & 2 & 0 & 1 & 2 & 0 & 0 & 0 & 2 & 0 & 0 \\
\hline $d$ & 1 & 0 & 2 & 2 & 0 & 0 & 0 & 1 & 0 & 2 & 2 & 0 & 0 & 0 & 0 \\
\hline $s$ & 0 & 0 & 0 & 0 & 2 & 0 & 2 & 0 & 0 & 0 & 0 & 2 & 0 & 2 & 2 \\
\hline $\overline{u}$ & 1 & 0 & 2 & 0 & 0 & 0 & 2 & 1 & 0 & 2 & 0 & 0 & 0 & 2 & 0 \\
\hline $\overline{d}$ & 1 & 2 & 0 & 0 & 2 & 0 & 0 & 1 & 2 & 0 & 0 & 2 & 0 & 0 & 0 \\
\hline $\overline{s}$ & 0 & 0 & 0 & 2 & 0 & 2 & 0 & 0 & 0 & 0 & 2 & 0 & 2 & 0 & 2 \\
\hline
\end{tabular}
}
\end{center}
\end{table}

\section{Vector Meson Splitting Functions}
\label{APP_VM_SPLITT}

 In this appendix we present the details for derivation of vector meson splitting functions. We start with the expression in the Eq.~(\ref{EQ_QUARK_FRAG_V}):
\begin{eqnarray}
\label{EQ_QUARK_FRAG_AP}
\nonumber
d_{q}^{m}(z)=    \frac{C_{q}^m}{2}  g_{mqQ}^{2} \frac{z}{2} \int \frac{ d^{2}\mathbf{k}_{T}}{(2\pi)^{3}} \frac{1}{2 p^- (1/z-1)} \frac{\mathrm{Tr}[(\slashed{k}+M_1)\gamma^{+}(\slashed{k}+M_1)  \gamma_{\mu} (\slashed{k}-\slashed{p}+M_{2}) \gamma_{\nu}] } {(k^2-M_1^2)^{2}}\\ 
 \times \left( -g^{\mu \nu}  +\frac{p^\mu p^\nu}{p^2} \right)|_{k_{-} = p_{-}/z,\ (k-p)^{2} =M_{2}^{2} } &,
\end{eqnarray}
where we used the following relation to sum over the vector meson polarizations:
\begin{eqnarray}
\label{EQ_POL_VEC_SUM}
 \sum_{\lambda=\mp1, 0}\epsilon^\mu(\lambda, p) \epsilon^{*\nu}(\lambda, p) = \left( -g^{\mu \nu} +\frac{p^\mu p^\nu}{p^2} \right).
\end{eqnarray}

We split the $\mathrm{Tr}$ in the Eq.~(\ref{EQ_QUARK_FRAG_AP}) into two parts:
\begin{align}
\label{EQ_FRAG_TR}
\mathrm{Tr^d_V} \equiv \mathrm{Tr}\left[(\slashed{k}+M_1)\gamma^{+}(\slashed{k}+M_1)\left(- \gamma^{\mu} (\slashed{k}-\slashed{p}+M_{2}) \gamma_{\mu} + \frac{\slashed{p} (\slashed{k}-\slashed{p}+M_{2}) \slashed{p} }{p^2} \right)\right ]=
\mathrm{Tr1^d_V}+\frac{\mathrm{Tr2^d_V}}{p^2}.
\end{align}
Here 
\begin{align}
\label{EQ_FRAG_TR1}
&\mathrm{Tr1^d_V} \equiv -\mathrm{Tr}\left[(\slashed{k}+M_1)\gamma^{+}(\slashed{k}+M_1)\gamma^{\mu} (\slashed{k}-\slashed{p}+M_{2}) \gamma_{\mu} \right ]&\\
& =8 \frac{p^-}{z(1-z)} \left \{z^2k_T^2 +(M_2-(1-z)M_1)^2  - 2(1-z)M_1 M_2 \right \}.&
\end{align}

And
\begin{align}
\label{EQ_FRAG_TR2}
&\mathrm{Tr2^d_V} \equiv \mathrm{Tr}\left[(\slashed{k}+M_1)\gamma^{+}(\slashed{k}+M_1) \slashed{p} (\slashed{k}-\slashed{p}+M_{2}) \slashed{p}\right ]&\\
&=4 \frac{p^-}{z(1-z) z^2}
\left\{ ((1-z) p^2+z^2 (k_T^2+M_1 M_2))^2+k_T^2 (M_1-M_2)^2 z^4 \right\} .&
\end{align}

In the above derivations, we used the on-mass-shell condition for the emitted vector meson with mass $m_m$ and fragmented quark in the frame where $\mathbf{p}_T=0$. Thus, we obtain for the elementary splitting function
 \begin{align}
\label{EQ_QUARK_FRAG_F}
&d_{q}^{m}(z)=\frac{C_{q}^m}{2}  g_{mqQ}^{2} z 
\int \frac{ d^{2}\mathbf{p}_\perp}{(2\pi)^{3}} \nonumber \\
 &\frac{
 2  \left(p_\perp^2 +((1-z)M_1-M_2)^2\right)
+\frac{1}{z^2 m_m^2}\left( \left(p_\perp^2 - z^2 M_1 M_2+(1-z) m_m^2\right)^2+p_\perp^2 z^2 \left(M_1+M_2\right)^2  \right)
 }
{ \left (p_\perp^2 + z(z-1)M_1^2+ z M_2^2 + (1-z)m_m^2  \right)^2}.
\end{align}

\section{Quark-Meson Couplings with Proper-Time Regularization}
\label{APP_QM_COUPLING}

 Here, we derive the vector meson-quark coupling constants in NJL model using the PT regularization scheme~\cite{Ebert:1996vx, Hellstern:1997nv, Bentz:2001vc} to calculate the quark--vector-meson couplings.
 \begin{eqnarray}
\label{EQ_PROP_T}
\frac{1}{X^n}= \frac{1}{(n-1)!}\int_0^{\infty} d\tau\ \tau^{n-1} e^{-\tau X} \rightarrow \frac{1}{(n-1)!}\int_{1/\Lambda_{UV}^2}^{1/\Lambda_{IR}^2} d\tau\ \tau^{n-1} e^{-\tau X},
\end{eqnarray}
where $X$ denotes the denominators of the loop integral after an appropriate momentum shift and Wick rotation,  $\Lambda_{IR}$ and $\Lambda_{UV}$ are the infrared (IR) and ultraviolet (UV) cut-offs, respectively.  It was demonstrated in Refs.~\cite{Ebert:1996vx, Hellstern:1997nv} that $\Lambda_{IR}$ mimics confinement by eliminating the unphysical thresholds for the decay of the hadrons into quarks, while in Ref.~\cite{Bentz:2001vc} it has been shown that this is crucial to describe saturation of the nuclear matter binding energy in the NJL model. Here we fix $M_u=0.3~\rm{GeV}$ as in our previous study \cite{Matevosyan:2010hh} and $\Lambda_{IR}= 200~\mathrm{MeV}$ as in previous model calculations of Ref.~\cite{Mineo:2003vc} and fit $\Lambda_{UV}$ to reproduce the experimental value of pion decay constant $f_\pi= 93~\mathrm{MeV}$. We then obtain the strange constituent quark mass requiring the calculated kaon mass to reproduce the experimentally measured value $m_K = 495~\mathrm{MeV}$. We will verify that the pseudoscalar meson-quark couplings are close to the values calculated in the LB regularization scheme, indicating that it is reasonable to use the PT scheme for the calculation of the vector meson-quark couplings, while keeping the LB scheme for the regularization of the elementary fragmentation functions.

 \subsection{Pseudoscalar Meson-Quark Coupling}

  Here we follow our previous work of  Ref.~\cite{Matevosyan:2010hh}. The quark-meson coupling constant is determined from the residue at the pole in the quark-antiquark t-matrix at the mass of the meson under consideration. This involves the derivative of the familiar quark-bubble graph
\begin{equation}
\label{EQ_VACUUM_BUBBLE}
\Pi(p^2)=2N_{c}i\int \frac{d^{4}k}{(2\pi)^{4}} Tr[\gamma_{5}S_{1}(k)\gamma_{5}S_{2}(k-p)],
\end{equation}

\begin{equation}
\label{EQ_COUPLING_KQQ}
\frac{1}{g^{2}_{mqQ}}=- \left( \frac{\partial \Pi(p^2)}{\partial p^{2}}  \right)_{p^{2}=m^{2}_{m}}.
\end{equation}
 Here $\mathrm{Tr}$ denotes the Dirac trace and the subscripts on the quark propagators denote quarks of different flavor - also indicated by $q$ and $Q$, where the meson of type $m$ under consideration has mass $m_m$ and flavor structure $m=q\overline{Q}$. Using the proper-time regularization of Eq.~(\ref{EQ_PROP_T}) we can calculate the above integral as
\begin{align}
\label{EQ_VACUUM_BUBBLE_TRANS}
\Pi(p^2)=&8N_{c}i\int \frac{d^{4}k}{(2\pi)^{4}} \frac{-k^2+k\cdot p + M_1 M_2}{(k^2-M_1^2)((k-p)^2-M_2^2))}\\
=&- \frac{N_{c}}{2\pi^2} \int_{1/\Lambda_{UV}^2}^{1/\Lambda_{IR}^2} \frac{d\tau}{\tau^2}\ \int_0^1 dx\left[1+\tau (B_{12}(x,p^2)-A_{12}(x,p^2))\right]\ e^{-\tau A_{12}(x,p^2)},
\end{align}
where 
\begin{align}
\label{EQ_A12_B12}
A_{12}(x,p^2)&\equiv (1-x)M_1^2 +xM_2^2 -x(1-x)p^2,\\
B_{12}(x,p^2)&\equiv  x(1-x)p^2+M_1 M_2.
\end{align}

For the quark-meson coupling we have:
\begin{eqnarray}
\label{EQ_COUPL_PROPT}
g_{mqQ}^{-2}= \frac{N_{c}}{2\pi^2} \int_{1/\Lambda_{UV}^2}^{1/\Lambda_{IR}^2} \frac{d\tau}{\tau}\ \int_0^1 dx\ x(1-x) \left[ 3+ \tau( B_{12}(x,m^2)
   - A_{12}(x,m^2)  )\right]\ e^{-\tau A_{12}(x,m^2)}.
\end{eqnarray}
  
\subsection{Decay Constant and Fixing the UV cutoff.}

 Here, we fix the $\Lambda_{UV}$ via the decay constant of the pion, which is obtained in the NJL model from the one loop contribution of the meson decay:
\begin{align}
\label{EQ_F_DEF}
&i p^{\mu}f_{m}= N_{c}g_{mqQ}\int \frac{d^{4}k}{(2\pi)^{4}} Tr[\gamma_{5}S_{1}(k)\gamma^{\mu}\gamma_{5}S_{2}(k-p)]|_{p^2=m_m^2},\\
&f_{m}=-\frac{ i N_{c}g_{mqQ}} {p^2}\int \frac{d^{4}k}{(2\pi)^{4}} Tr[\gamma_{5}S_{1}(k)\slashed{p}\gamma_{5}S_{2}(k-p)]|_{p^2=m_m^2}\\
&=\frac{ N_{c}g_{mqQ}} {4\pi^2} \int_0^1 dx\ ((1-x)M_1+x M_2)\ \left(Ei\left[-\frac{A_{12}(x,m_m^2)}{\Lambda^2_{IR}}\right] -Ei\left[-\frac{A_{12}(x,m_m^2)}{\Lambda^2_{UV}}\right]\right).
\end{align}

 Using the previously set values of $M_u=0.3~\rm{GeV}$ and $\Lambda_{IR}=0.2~\rm{GeV}$, the fit to the experimental value of $f_\pi=0.093~\rm{GeV}$ yields for the UV cutoff $\Lambda_{UV}=0.703~\rm{GeV}$. This allows us to determine the strange constituent quark mass, $M_s$, requiring that the kaon mass determined from the quark anti-quark t-matrix pole reproduces the experimental value. This gives $M_s=0.539~\rm{GeV}$. The value of $M_s$ in the PT scheme is only marginally larger than the value used in the previous study~\cite{Matevosyan:2010hh}, which was determined using LB regularization: $M_s^{(LB)}=0.537~\rm{GeV}$. Then we can confirm that  the values of the pseudoscalar meson-quark coupling constants calculated in both regularization schemes are indeed very close to each other
\begin{align}
\label{EQ_GS_COMP}
&g^{(LB)}_{\pi qQ} = 3.15,\ g^{(LB)}_{K qQ} = 3.39,\nonumber \\
&g^{(PT)}_{\pi qQ} = 3.16,\ g^{(PT)}_{K qQ} = 3.41.
\end{align}

\subsection{Vector Meson-Quark Couplings}
\label{SUB_SEC_VM_COUPL}
 The vector meson-quark coupling constant, $g_{mqQ}$, can be calculated in the familiar manner. First, we evaluate the quark-bubble graph in vector channel using the PT regularization scheme; here only the transverse component contributes to the corresponding t-matrix for our model Lagrangian with only the four-point quark-quark interactions \cite{Klimt:1989pm} 
\begin{align}
\label{EQ_VACUUM_BUBBLE_VEC_ALT}
 &\Pi_{VV}^{(T)}(p^2)\left( g^{\mu \nu} - \frac{p^\mu p^\nu}{p^2}\right)=2N_{c}i\int \frac{d^{4}k}{(2\pi)^{4}} Tr\left[\left( g^{\mu \alpha} - \frac{p^\mu p^\alpha}{p^2}\right)\gamma_{\alpha}S_{1}(k)\left( g^{\nu \beta} - \frac{p^\nu p^\beta}{p^2}\right)\gamma_{\beta}S_{2}(k-p)\right]\nonumber \\
&= - \left( g^{\mu \nu} - \frac{p^\mu p^\nu}{p^2}\right) \frac{N_{c}}{2\pi^2} \int_0^1 dx\ \int \frac{d \tau}{\tau}  \left(M_1 M_2 - (1-x)M_1^2 -xM_2^2 +2x(1-x) p^2 \right)e^{-\tau A_{12}(x,p^2)}\ .
\end{align}

 Thus we finally obtain for the coupling constant:
\begin{align}
\label{EQ_VEC_COUPL}
 &g^{-2}_{mqQ}=-\frac{\partial\Pi_{VV}^{(T)}(p^2)}{\partial p^2}|_{p^2=m_m^2}
\\
&=  \frac{N_{c}}{2\pi^2} \int_0^1 dx\ x(1-x)\int_{1/\Lambda^2_{UV}}^{1/\Lambda^2_{IR}}d \tau \left( M_1 M_2 - (1-x)M_1^2 -xM_2^2+2 x(1-x) m_m^2 +  \frac{2}{\tau}\right) e^{-\tau A_{12}(x,m_m^2)}. \nonumber
\end{align}

 The masses of the vector mesons can be determined within the NJL framework by fixing the quark coupling in the Lagrangian using the experimental value for the $\rho$ meson. Then the masses for the $K^*$ and $\phi$ mesons are
\begin{equation}
\label{EQ_VEC_MASS}
m_{K^*}^{(NJL)}=0.936~\mathrm{GeV},\ \ m_\phi^{(NJL)} =1.088~\mathrm{GeV},
\end{equation}
which are reasonably close to the experimental values, given the simplistic Lagrangian used. In this work we choose to use the experimental values for all the vector mesons in calculating the couplings and the splitting functions in an attempt to best describe the measured fragmentation functions. Then the corresponding values of the couplings for the vector mesons considered in this article are
 \begin{equation}
\label{EQ_VEC_COUPL_SAMP}
 g_{\rho qQ}=1.5,~~
 g_{K^{*} qQ}=1.91,~~ 
 g_{\phi qQ}=2.29.
\end{equation}
 
\bibliographystyle{apsrev}
\bibliography{../Bibliography/fragment}

\begin{thebibliography}{52}
\expandafter\ifx\csname natexlab\endcsname\relax\def\natexlab#1{#1}\fi
\expandafter\ifx\csname bibnamefont\endcsname\relax
  \def\bibnamefont#1{#1}\fi
\expandafter\ifx\csname bibfnamefont\endcsname\relax
  \def\bibfnamefont#1{#1}\fi
\expandafter\ifx\csname citenamefont\endcsname\relax
  \def\citenamefont#1{#1}\fi
\expandafter\ifx\csname url\endcsname\relax
  \def\url#1{\texttt{#1}}\fi
\expandafter\ifx\csname urlprefix\endcsname\relax\def\urlprefix{URL }\fi
\providecommand{\bibinfo}[2]{#2}
\providecommand{\eprint}[2][]{\url{#2}}

\bibitem[{\citenamefont{Accardi et~al.}(2010)\citenamefont{Accardi, Arleo,
  Brooks, D'Enterria, and Muccifora}}]{Accardi:2009qv}
\bibinfo{author}{\bibfnamefont{A.}~\bibnamefont{Accardi}},
  \bibinfo{author}{\bibfnamefont{F.}~\bibnamefont{Arleo}},
  \bibinfo{author}{\bibfnamefont{W.~K.} \bibnamefont{Brooks}},
  \bibinfo{author}{\bibfnamefont{D.}~\bibnamefont{D'Enterria}},
  \bibnamefont{and}
  \bibinfo{author}{\bibfnamefont{V.}~\bibnamefont{Muccifora}},
  \bibinfo{journal}{Riv.Nuovo Cim.} \textbf{\bibinfo{volume}{32}},
  \bibinfo{pages}{439} (\bibinfo{year}{2010}), \eprint{0907.3534}.

\bibitem[{\citenamefont{Buskulic et~al.}(1995)}]{Buskulic:1994ft}
\bibinfo{author}{\bibfnamefont{D.}~\bibnamefont{Buskulic}} \bibnamefont{et~al.}
  (\bibinfo{collaboration}{ALEPH Collaboration}), \bibinfo{journal}{Z.Phys.}
  \textbf{\bibinfo{volume}{C66}}, \bibinfo{pages}{355} (\bibinfo{year}{1995}).

\bibitem[{\citenamefont{Abreu et~al.}(1998)}]{Abreu:1998vq}
\bibinfo{author}{\bibfnamefont{P.}~\bibnamefont{Abreu}} \bibnamefont{et~al.}
  (\bibinfo{collaboration}{DELPHI Collaboration}),
  \bibinfo{journal}{Eur.Phys.J.} \textbf{\bibinfo{volume}{C5}},
  \bibinfo{pages}{585} (\bibinfo{year}{1998}).

\bibitem[{\citenamefont{Abbiendi et~al.}(2000)}]{Abbiendi:1999ry}
\bibinfo{author}{\bibfnamefont{G.}~\bibnamefont{Abbiendi}} \bibnamefont{et~al.}
  (\bibinfo{collaboration}{OPAL Collaboration}), \bibinfo{journal}{Eur.Phys.J.}
  \textbf{\bibinfo{volume}{C16}}, \bibinfo{pages}{407} (\bibinfo{year}{2000}),
  \eprint{hep-ex/0001054}.

\bibitem[{\citenamefont{Aihara et~al.}(1987)}]{Aihara:1986mv}
\bibinfo{author}{\bibfnamefont{H.}~\bibnamefont{Aihara}} \bibnamefont{et~al.}
  (\bibinfo{collaboration}{TPC/Two Gamma Collaboration}),
  \bibinfo{journal}{Phys.Lett.} \textbf{\bibinfo{volume}{B184}},
  \bibinfo{pages}{299} (\bibinfo{year}{1987}).

\bibitem[{\citenamefont{Abe et~al.}(1999)}]{Abe:1998zs}
\bibinfo{author}{\bibfnamefont{K.}~\bibnamefont{Abe}} \bibnamefont{et~al.}
  (\bibinfo{collaboration}{SLD Collaboration}), \bibinfo{journal}{Phys.Rev.}
  \textbf{\bibinfo{volume}{D59}}, \bibinfo{pages}{052001}
  (\bibinfo{year}{1999}), \eprint{hep-ex/9805029}.

\bibitem[{\citenamefont{Adler et~al.}(2003)}]{Adler:2003pb}
\bibinfo{author}{\bibfnamefont{S.~S.} \bibnamefont{Adler}} \bibnamefont{et~al.}
  (\bibinfo{collaboration}{PHENIX Collaboration}),
  \bibinfo{journal}{Phys.Rev.Lett.} \textbf{\bibinfo{volume}{91}},
  \bibinfo{pages}{241803} (\bibinfo{year}{2003}), \eprint{hep-ex/0304038}.

\bibitem[{\citenamefont{Arsene et~al.}(2007)}]{Arsene:2007jd}
\bibinfo{author}{\bibfnamefont{I.}~\bibnamefont{Arsene}} \bibnamefont{et~al.}
  (\bibinfo{collaboration}{BRAHMS Collaboration}),
  \bibinfo{journal}{Phys.Rev.Lett.} \textbf{\bibinfo{volume}{98}},
  \bibinfo{pages}{252001} (\bibinfo{year}{2007}), \eprint{hep-ex/0701041}.

\bibitem[{\citenamefont{Abelev et~al.}(2007)}]{Abelev:2006cs}
\bibinfo{author}{\bibfnamefont{B.}~\bibnamefont{Abelev}} \bibnamefont{et~al.}
  (\bibinfo{collaboration}{STAR Collaboration}), \bibinfo{journal}{Phys.Rev.}
  \textbf{\bibinfo{volume}{C75}}, \bibinfo{pages}{064901}
  (\bibinfo{year}{2007}), \eprint{nucl-ex/0607033}.

\bibitem[{\citenamefont{de~Florian
  et~al.}(2007{\natexlab{a}})\citenamefont{de~Florian, Sassot, and
  Stratmann}}]{deFlorian:2007aj}
\bibinfo{author}{\bibfnamefont{D.}~\bibnamefont{de~Florian}},
  \bibinfo{author}{\bibfnamefont{R.}~\bibnamefont{Sassot}}, \bibnamefont{and}
  \bibinfo{author}{\bibfnamefont{M.}~\bibnamefont{Stratmann}},
  \bibinfo{journal}{Phys. Rev.} \textbf{\bibinfo{volume}{D75}},
  \bibinfo{pages}{114010} (\bibinfo{year}{2007}{\natexlab{a}}),
  \eprint{hep-ph/0703242}.

\bibitem[{\citenamefont{de~Florian
  et~al.}(2007{\natexlab{b}})\citenamefont{de~Florian, Sassot, and
  Stratmann}}]{deFlorian:2007hc}
\bibinfo{author}{\bibfnamefont{D.}~\bibnamefont{de~Florian}},
  \bibinfo{author}{\bibfnamefont{R.}~\bibnamefont{Sassot}}, \bibnamefont{and}
  \bibinfo{author}{\bibfnamefont{M.}~\bibnamefont{Stratmann}},
  \bibinfo{journal}{Phys.Rev.} \textbf{\bibinfo{volume}{D76}},
  \bibinfo{pages}{074033} (\bibinfo{year}{2007}{\natexlab{b}}),
  \eprint{0707.1506}.

\bibitem[{\citenamefont{Hirai et~al.}(2007)\citenamefont{Hirai, Kumano, Nagai,
  and Sudoh}}]{Hirai:2007cx}
\bibinfo{author}{\bibfnamefont{M.}~\bibnamefont{Hirai}},
  \bibinfo{author}{\bibfnamefont{S.}~\bibnamefont{Kumano}},
  \bibinfo{author}{\bibfnamefont{T.~H.} \bibnamefont{Nagai}}, \bibnamefont{and}
  \bibinfo{author}{\bibfnamefont{K.}~\bibnamefont{Sudoh}},
  \bibinfo{journal}{Phys. Rev.} \textbf{\bibinfo{volume}{D75}},
  \bibinfo{pages}{094009} (\bibinfo{year}{2007}), \eprint{hep-ph/0702250}.

\bibitem[{\citenamefont{Airapetian et~al.}(2007)}]{Airapetian:2007vu}
\bibinfo{author}{\bibfnamefont{A.}~\bibnamefont{Airapetian}}
  \bibnamefont{et~al.} (\bibinfo{collaboration}{HERMES Collaboration}),
  \bibinfo{journal}{Nucl.Phys.} \textbf{\bibinfo{volume}{B780}},
  \bibinfo{pages}{1} (\bibinfo{year}{2007}), \eprint{0704.3270}.

\bibitem[{\citenamefont{Avakian et~al.}(2010)}]{Avakian:2010ae}
\bibinfo{author}{\bibfnamefont{H.}~\bibnamefont{Avakian}} \bibnamefont{et~al.}
  (\bibinfo{collaboration}{The CLAS}), \bibinfo{journal}{Phys. Rev. Lett.}
  \textbf{\bibinfo{volume}{105}}, \bibinfo{pages}{262002}
  (\bibinfo{year}{2010}), \eprint{1003.4549}.

\bibitem[{\citenamefont{Avakian et~al.}(2005)\citenamefont{Avakian, Bosted,
  Burkert, and Elouadrhiri}}]{Avakian:2005ps}
\bibinfo{author}{\bibfnamefont{H.}~\bibnamefont{Avakian}},
  \bibinfo{author}{\bibfnamefont{P.~E.} \bibnamefont{Bosted}},
  \bibinfo{author}{\bibfnamefont{V.}~\bibnamefont{Burkert}}, \bibnamefont{and}
  \bibinfo{author}{\bibfnamefont{L.}~\bibnamefont{Elouadrhiri}}
  (\bibinfo{collaboration}{CLAS Collaboration}), \bibinfo{journal}{AIP
  Conf.Proc.} \textbf{\bibinfo{volume}{792}}, \bibinfo{pages}{945}
  (\bibinfo{year}{2005}), \eprint{nucl-ex/0509032}.

\bibitem[{\citenamefont{Mkrtchyan et~al.}(2008)\citenamefont{Mkrtchyan, Bosted,
  Adams, Ahmidouch, Angelescu et~al.}}]{Mkrtchyan:2007sr}
\bibinfo{author}{\bibfnamefont{H.}~\bibnamefont{Mkrtchyan}},
  \bibinfo{author}{\bibfnamefont{P.}~\bibnamefont{Bosted}},
  \bibinfo{author}{\bibfnamefont{G.}~\bibnamefont{Adams}},
  \bibinfo{author}{\bibfnamefont{A.}~\bibnamefont{Ahmidouch}},
  \bibinfo{author}{\bibfnamefont{T.}~\bibnamefont{Angelescu}},
  \bibnamefont{et~al.}, \bibinfo{journal}{Phys.Lett.}
  \textbf{\bibinfo{volume}{B665}}, \bibinfo{pages}{20} (\bibinfo{year}{2008}),
  \eprint{0709.3020}.

\bibitem[{\citenamefont{Brooks and Hakobyan}(2009)}]{Brooks:2009xg}
\bibinfo{author}{\bibfnamefont{W.}~\bibnamefont{Brooks}} \bibnamefont{and}
  \bibinfo{author}{\bibfnamefont{H.}~\bibnamefont{Hakobyan}},
  \bibinfo{journal}{Nucl.Phys.} \textbf{\bibinfo{volume}{A830}},
  \bibinfo{pages}{361C} (\bibinfo{year}{2009}), \eprint{0907.4606}.

\bibitem[{\citenamefont{Brooks}(2010)}]{Brooks:2010rz}
\bibinfo{author}{\bibfnamefont{W.}~\bibnamefont{Brooks}},
  \bibinfo{journal}{PoS} \textbf{\bibinfo{volume}{DIS2010}},
  \bibinfo{pages}{258} (\bibinfo{year}{2010}), \eprint{1008.0131}.

\bibitem[{\citenamefont{Thomas}(2009)}]{Thomas:2009ei}
\bibinfo{author}{\bibfnamefont{A.~W.} \bibnamefont{Thomas}}
  (\bibinfo{year}{2009}), \eprint{0907.4785}.

\bibitem[{\citenamefont{Nambu and
  Jona-Lasinio}(1961{\natexlab{a}})}]{Nambu:1961tp}
\bibinfo{author}{\bibfnamefont{Y.}~\bibnamefont{Nambu}} \bibnamefont{and}
  \bibinfo{author}{\bibfnamefont{G.}~\bibnamefont{Jona-Lasinio}},
  \bibinfo{journal}{Phys. Rev.} \textbf{\bibinfo{volume}{122}},
  \bibinfo{pages}{345} (\bibinfo{year}{1961}{\natexlab{a}}).

\bibitem[{\citenamefont{Nambu and
  Jona-Lasinio}(1961{\natexlab{b}})}]{Nambu:1961fr}
\bibinfo{author}{\bibfnamefont{Y.}~\bibnamefont{Nambu}} \bibnamefont{and}
  \bibinfo{author}{\bibfnamefont{G.}~\bibnamefont{Jona-Lasinio}},
  \bibinfo{journal}{Phys. Rev.} \textbf{\bibinfo{volume}{124}},
  \bibinfo{pages}{246} (\bibinfo{year}{1961}{\natexlab{b}}).

\bibitem[{\citenamefont{Cloet et~al.}(2005)\citenamefont{Cloet, Bentz, and
  Thomas}}]{Cloet:2005pp}
\bibinfo{author}{\bibfnamefont{I.~C.} \bibnamefont{Cloet}},
  \bibinfo{author}{\bibfnamefont{W.}~\bibnamefont{Bentz}}, \bibnamefont{and}
  \bibinfo{author}{\bibfnamefont{A.~W.} \bibnamefont{Thomas}},
  \bibinfo{journal}{Phys. Lett.} \textbf{\bibinfo{volume}{B621}},
  \bibinfo{pages}{246} (\bibinfo{year}{2005}), \eprint{hep-ph/0504229}.

\bibitem[{\citenamefont{Field and Feynman}(1978)}]{Field:1977fa}
\bibinfo{author}{\bibfnamefont{R.~D.} \bibnamefont{Field}} \bibnamefont{and}
  \bibinfo{author}{\bibfnamefont{R.~P.} \bibnamefont{Feynman}},
  \bibinfo{journal}{Nucl. Phys.} \textbf{\bibinfo{volume}{B136}},
  \bibinfo{pages}{1} (\bibinfo{year}{1978}).

\bibitem[{\citenamefont{Ito et~al.}(2009)\citenamefont{Ito, Bentz, Cloet,
  Thomas, and Yazaki}}]{Ito:2009zc}
\bibinfo{author}{\bibfnamefont{T.}~\bibnamefont{Ito}},
  \bibinfo{author}{\bibfnamefont{W.}~\bibnamefont{Bentz}},
  \bibinfo{author}{\bibfnamefont{I.~C.} \bibnamefont{Cloet}},
  \bibinfo{author}{\bibfnamefont{A.~W.} \bibnamefont{Thomas}},
  \bibnamefont{and} \bibinfo{author}{\bibfnamefont{K.}~\bibnamefont{Yazaki}},
  \bibinfo{journal}{Phys. Rev.} \textbf{\bibinfo{volume}{D80}},
  \bibinfo{pages}{074008} (\bibinfo{year}{2009}), \eprint{0906.5362}.

\bibitem[{\citenamefont{Matevosyan et~al.}(2011)\citenamefont{Matevosyan,
  Thomas, and Bentz}}]{Matevosyan:2010hh}
\bibinfo{author}{\bibfnamefont{H.~H.} \bibnamefont{Matevosyan}},
  \bibinfo{author}{\bibfnamefont{A.~W.} \bibnamefont{Thomas}},
  \bibnamefont{and} \bibinfo{author}{\bibfnamefont{W.}~\bibnamefont{Bentz}},
  \bibinfo{journal}{Phys.Rev.} \textbf{\bibinfo{volume}{D83}},
  \bibinfo{pages}{074003} (\bibinfo{year}{2011}), \eprint{1011.1052}.

\bibitem[{\citenamefont{Mineo et~al.}(1999)\citenamefont{Mineo, Bentz, and
  Yazaki}}]{Mineo:1999eq}
\bibinfo{author}{\bibfnamefont{H.}~\bibnamefont{Mineo}},
  \bibinfo{author}{\bibfnamefont{W.}~\bibnamefont{Bentz}}, \bibnamefont{and}
  \bibinfo{author}{\bibfnamefont{K.}~\bibnamefont{Yazaki}},
  \bibinfo{journal}{Phys.Rev.} \textbf{\bibinfo{volume}{C60}},
  \bibinfo{pages}{065201} (\bibinfo{year}{1999}), \eprint{nucl-th/9907043}.

\bibitem[{\citenamefont{Jakob et~al.}(1997)\citenamefont{Jakob, Mulders, and
  Rodrigues}}]{Jakob:1997wg}
\bibinfo{author}{\bibfnamefont{R.}~\bibnamefont{Jakob}},
  \bibinfo{author}{\bibfnamefont{P.~J.} \bibnamefont{Mulders}},
  \bibnamefont{and}
  \bibinfo{author}{\bibfnamefont{J.}~\bibnamefont{Rodrigues}},
  \bibinfo{journal}{Nucl. Phys.} \textbf{\bibinfo{volume}{A626}},
  \bibinfo{pages}{937} (\bibinfo{year}{1997}), \eprint{hep-ph/9704335}.

\bibitem[{\citenamefont{Bacchetta et~al.}(2004)\citenamefont{Bacchetta,
  Schaefer, and Yang}}]{Bacchetta:2003rz}
\bibinfo{author}{\bibfnamefont{A.}~\bibnamefont{Bacchetta}},
  \bibinfo{author}{\bibfnamefont{A.}~\bibnamefont{Schaefer}}, \bibnamefont{and}
  \bibinfo{author}{\bibfnamefont{J.-J.} \bibnamefont{Yang}},
  \bibinfo{journal}{Phys. Lett.} \textbf{\bibinfo{volume}{B578}},
  \bibinfo{pages}{109} (\bibinfo{year}{2004}), \eprint{hep-ph/0309246}.

\bibitem[{\citenamefont{Bacchetta et~al.}(2008)\citenamefont{Bacchetta, Conti,
  and Radici}}]{Bacchetta:2008af}
\bibinfo{author}{\bibfnamefont{A.}~\bibnamefont{Bacchetta}},
  \bibinfo{author}{\bibfnamefont{F.}~\bibnamefont{Conti}}, \bibnamefont{and}
  \bibinfo{author}{\bibfnamefont{M.}~\bibnamefont{Radici}},
  \bibinfo{journal}{Phys. Rev.} \textbf{\bibinfo{volume}{D78}},
  \bibinfo{pages}{074010} (\bibinfo{year}{2008}), \eprint{0807.0323}.

\bibitem[{\citenamefont{Kitagawa and Sakemi}(2001)}]{Kitagawa:2001ig}
\bibinfo{author}{\bibfnamefont{H.}~\bibnamefont{Kitagawa}} \bibnamefont{and}
  \bibinfo{author}{\bibfnamefont{Y.}~\bibnamefont{Sakemi}},
  \bibinfo{journal}{Prog.Theor.Phys.} \textbf{\bibinfo{volume}{105}},
  \bibinfo{pages}{751} (\bibinfo{year}{2001}).

\bibitem[{\citenamefont{Yang}(2002)}]{Yang:2002gh}
\bibinfo{author}{\bibfnamefont{J.-J.} \bibnamefont{Yang}},
  \bibinfo{journal}{Phys. Rev.} \textbf{\bibinfo{volume}{D65}},
  \bibinfo{pages}{094035} (\bibinfo{year}{2002}).

\bibitem[{\citenamefont{Kato et~al.}(1993)\citenamefont{Kato, Bentz, Yazaki,
  and Tanaka}}]{Kato:1993zw}
\bibinfo{author}{\bibfnamefont{M.}~\bibnamefont{Kato}},
  \bibinfo{author}{\bibfnamefont{W.}~\bibnamefont{Bentz}},
  \bibinfo{author}{\bibfnamefont{K.}~\bibnamefont{Yazaki}}, \bibnamefont{and}
  \bibinfo{author}{\bibfnamefont{K.}~\bibnamefont{Tanaka}},
  \bibinfo{journal}{Nucl. Phys.} \textbf{\bibinfo{volume}{A551}},
  \bibinfo{pages}{541} (\bibinfo{year}{1993}).

\bibitem[{\citenamefont{Klimt et~al.}(1990)\citenamefont{Klimt, Lutz, Vogl, and
  Weise}}]{Klimt:1989pm}
\bibinfo{author}{\bibfnamefont{S.}~\bibnamefont{Klimt}},
  \bibinfo{author}{\bibfnamefont{M.}~\bibnamefont{Lutz}},
  \bibinfo{author}{\bibfnamefont{U.}~\bibnamefont{Vogl}}, \bibnamefont{and}
  \bibinfo{author}{\bibfnamefont{W.}~\bibnamefont{Weise}},
  \bibinfo{journal}{Nucl. Phys.} \textbf{\bibinfo{volume}{A516}},
  \bibinfo{pages}{429} (\bibinfo{year}{1990}).

\bibitem[{\citenamefont{Klevansky}(1992)}]{Klevansky:1992qe}
\bibinfo{author}{\bibfnamefont{S.~P.} \bibnamefont{Klevansky}},
  \bibinfo{journal}{Rev. Mod. Phys.} \textbf{\bibinfo{volume}{64}},
  \bibinfo{pages}{649} (\bibinfo{year}{1992}).

\bibitem[{\citenamefont{Bentz et~al.}(1999)\citenamefont{Bentz, Hama, Matsuki,
  and Yazaki}}]{Bentz:1999gx}
\bibinfo{author}{\bibfnamefont{W.}~\bibnamefont{Bentz}},
  \bibinfo{author}{\bibfnamefont{T.}~\bibnamefont{Hama}},
  \bibinfo{author}{\bibfnamefont{T.}~\bibnamefont{Matsuki}}, \bibnamefont{and}
  \bibinfo{author}{\bibfnamefont{K.}~\bibnamefont{Yazaki}},
  \bibinfo{journal}{Nucl. Phys.} \textbf{\bibinfo{volume}{A651}},
  \bibinfo{pages}{143} (\bibinfo{year}{1999}), \eprint{hep-ph/9901377}.

\bibitem[{\citenamefont{Chliapnikov}(1999)}]{Chliapnikov:1999qi}
\bibinfo{author}{\bibfnamefont{P.~V.} \bibnamefont{Chliapnikov}},
  \bibinfo{journal}{Phys. Lett.} \textbf{\bibinfo{volume}{B462}},
  \bibinfo{pages}{341} (\bibinfo{year}{1999}).

\bibitem[{\citenamefont{Ebert et~al.}(1996)\citenamefont{Ebert, Feldmann, and
  Reinhardt}}]{Ebert:1996vx}
\bibinfo{author}{\bibfnamefont{D.}~\bibnamefont{Ebert}},
  \bibinfo{author}{\bibfnamefont{T.}~\bibnamefont{Feldmann}}, \bibnamefont{and}
  \bibinfo{author}{\bibfnamefont{H.}~\bibnamefont{Reinhardt}},
  \bibinfo{journal}{Phys. Lett.} \textbf{\bibinfo{volume}{B388}},
  \bibinfo{pages}{154} (\bibinfo{year}{1996}), \eprint{hep-ph/9608223}.

\bibitem[{\citenamefont{Hellstern et~al.}(1997)\citenamefont{Hellstern,
  Alkofer, and Reinhardt}}]{Hellstern:1997nv}
\bibinfo{author}{\bibfnamefont{G.}~\bibnamefont{Hellstern}},
  \bibinfo{author}{\bibfnamefont{R.}~\bibnamefont{Alkofer}}, \bibnamefont{and}
  \bibinfo{author}{\bibfnamefont{H.}~\bibnamefont{Reinhardt}},
  \bibinfo{journal}{Nucl. Phys.} \textbf{\bibinfo{volume}{A625}},
  \bibinfo{pages}{697} (\bibinfo{year}{1997}), \eprint{hep-ph/9706551}.

\bibitem[{\citenamefont{Bentz and Thomas}(2001)}]{Bentz:2001vc}
\bibinfo{author}{\bibfnamefont{W.}~\bibnamefont{Bentz}} \bibnamefont{and}
  \bibinfo{author}{\bibfnamefont{A.~W.} \bibnamefont{Thomas}},
  \bibinfo{journal}{Nucl. Phys.} \textbf{\bibinfo{volume}{A696}},
  \bibinfo{pages}{138} (\bibinfo{year}{2001}), \eprint{nucl-th/0105022}.

\bibitem[{\citenamefont{Drell et~al.}(1969)\citenamefont{Drell, Levy, and
  Yan}}]{Drell:1969jm}
\bibinfo{author}{\bibfnamefont{S.~D.} \bibnamefont{Drell}},
  \bibinfo{author}{\bibfnamefont{D.~J.} \bibnamefont{Levy}}, \bibnamefont{and}
  \bibinfo{author}{\bibfnamefont{T.-M.} \bibnamefont{Yan}},
  \bibinfo{journal}{Phys. Rev.} \textbf{\bibinfo{volume}{187}},
  \bibinfo{pages}{2159} (\bibinfo{year}{1969}).

\bibitem[{\citenamefont{Blumlein et~al.}(2000)\citenamefont{Blumlein,
  Ravindran, and van Neerven}}]{Blumlein:2000wh}
\bibinfo{author}{\bibfnamefont{J.}~\bibnamefont{Blumlein}},
  \bibinfo{author}{\bibfnamefont{V.}~\bibnamefont{Ravindran}},
  \bibnamefont{and} \bibinfo{author}{\bibfnamefont{W.~L.} \bibnamefont{van
  Neerven}}, \bibinfo{journal}{Nucl. Phys.} \textbf{\bibinfo{volume}{B586}},
  \bibinfo{pages}{349} (\bibinfo{year}{2000}), \eprint{hep-ph/0004172}.

\bibitem[{\citenamefont{Mineo et~al.}(2004)\citenamefont{Mineo, Bentz, Ishii,
  Thomas, and Yazaki}}]{Mineo:2003vc}
\bibinfo{author}{\bibfnamefont{H.}~\bibnamefont{Mineo}},
  \bibinfo{author}{\bibfnamefont{W.}~\bibnamefont{Bentz}},
  \bibinfo{author}{\bibfnamefont{N.}~\bibnamefont{Ishii}},
  \bibinfo{author}{\bibfnamefont{A.~W.} \bibnamefont{Thomas}},
  \bibnamefont{and} \bibinfo{author}{\bibfnamefont{K.}~\bibnamefont{Yazaki}},
  \bibinfo{journal}{Nucl. Phys.} \textbf{\bibinfo{volume}{A735}},
  \bibinfo{pages}{482} (\bibinfo{year}{2004}), \eprint{nucl-th/0312097}.

\bibitem[{\citenamefont{Andersson et~al.}(1983)\citenamefont{Andersson,
  Gustafson, Ingelman, and Sjostrand}}]{Andersson:1983ia}
\bibinfo{author}{\bibfnamefont{B.}~\bibnamefont{Andersson}},
  \bibinfo{author}{\bibfnamefont{G.}~\bibnamefont{Gustafson}},
  \bibinfo{author}{\bibfnamefont{G.}~\bibnamefont{Ingelman}}, \bibnamefont{and}
  \bibinfo{author}{\bibfnamefont{T.}~\bibnamefont{Sjostrand}},
  \bibinfo{journal}{Phys. Rept.} \textbf{\bibinfo{volume}{97}},
  \bibinfo{pages}{31} (\bibinfo{year}{1983}).

\bibitem[{\citenamefont{Sjostrand}(1982)}]{Sjostrand:1982fn}
\bibinfo{author}{\bibfnamefont{T.}~\bibnamefont{Sjostrand}},
  \bibinfo{journal}{Comput. Phys. Commun.} \textbf{\bibinfo{volume}{27}},
  \bibinfo{pages}{243} (\bibinfo{year}{1982}).

\bibitem[{\citenamefont{Sjostrand et~al.}(2008)\citenamefont{Sjostrand, Mrenna,
  and Skands}}]{Sjostrand:2007gs}
\bibinfo{author}{\bibfnamefont{T.}~\bibnamefont{Sjostrand}},
  \bibinfo{author}{\bibfnamefont{S.}~\bibnamefont{Mrenna}}, \bibnamefont{and}
  \bibinfo{author}{\bibfnamefont{P.~Z.} \bibnamefont{Skands}},
  \bibinfo{journal}{Comput. Phys. Commun.} \textbf{\bibinfo{volume}{178}},
  \bibinfo{pages}{852} (\bibinfo{year}{2008}), \eprint{0710.3820}.

\bibitem[{\citenamefont{Ritter and Ranft}(1980)}]{Ritter:1979mk}
\bibinfo{author}{\bibfnamefont{S.}~\bibnamefont{Ritter}} \bibnamefont{and}
  \bibinfo{author}{\bibfnamefont{J.}~\bibnamefont{Ranft}},
  \bibinfo{journal}{Acta Phys.Polon.} \textbf{\bibinfo{volume}{B11}},
  \bibinfo{pages}{259} (\bibinfo{year}{1980}).

\bibitem[{\citenamefont{Nakamura et~al.}(2010)}]{Nakamura:2010zzi}
\bibinfo{author}{\bibfnamefont{K.}~\bibnamefont{Nakamura}} \bibnamefont{et~al.}
  (\bibinfo{collaboration}{Particle Data Group}), \bibinfo{journal}{J. Phys.}
  \textbf{\bibinfo{volume}{G37}}, \bibinfo{pages}{075021}
  (\bibinfo{year}{2010}).

\bibitem[{\citenamefont{Matsuyama et~al.}(2007)\citenamefont{Matsuyama, Sato,
  and Lee}}]{Matsuyama:2006rp}
\bibinfo{author}{\bibfnamefont{A.}~\bibnamefont{Matsuyama}},
  \bibinfo{author}{\bibfnamefont{T.}~\bibnamefont{Sato}}, \bibnamefont{and}
  \bibinfo{author}{\bibfnamefont{T.~S.~H.} \bibnamefont{Lee}},
  \bibinfo{journal}{Phys. Rept.} \textbf{\bibinfo{volume}{439}},
  \bibinfo{pages}{193} (\bibinfo{year}{2007}), \eprint{nucl-th/0608051}.

\bibitem[{\citenamefont{Andersson et~al.}(1978)\citenamefont{Andersson,
  Gustafson, and Peterson}}]{Andersson:1977xs}
\bibinfo{author}{\bibfnamefont{B.}~\bibnamefont{Andersson}},
  \bibinfo{author}{\bibfnamefont{G.}~\bibnamefont{Gustafson}},
  \bibnamefont{and} \bibinfo{author}{\bibfnamefont{C.}~\bibnamefont{Peterson}},
  \bibinfo{journal}{Nucl. Phys.} \textbf{\bibinfo{volume}{B135}},
  \bibinfo{pages}{273} (\bibinfo{year}{1978}).

\bibitem[{\citenamefont{Cloet et~al.}(2008)\citenamefont{Cloet, Bentz, and
  Thomas}}]{Cloet:2007em}
\bibinfo{author}{\bibfnamefont{I.~C.} \bibnamefont{Cloet}},
  \bibinfo{author}{\bibfnamefont{W.}~\bibnamefont{Bentz}}, \bibnamefont{and}
  \bibinfo{author}{\bibfnamefont{A.~W.} \bibnamefont{Thomas}},
  \bibinfo{journal}{Phys. Lett.} \textbf{\bibinfo{volume}{B659}},
  \bibinfo{pages}{214} (\bibinfo{year}{2008}), \eprint{0708.3246}.

\bibitem[{\citenamefont{Botje}(2011)}]{Botje:2010ay}
\bibinfo{author}{\bibfnamefont{M.}~\bibnamefont{Botje}},
  \bibinfo{journal}{Comput.Phys.Commun.} \textbf{\bibinfo{volume}{182}},
  \bibinfo{pages}{490} (\bibinfo{year}{2011}), \eprint{1005.1481}.

\bibitem[{\citenamefont{Cloet et~al.}(2006)\citenamefont{Cloet, Bentz, and
  Thomas}}]{Cloet:2006bq}
\bibinfo{author}{\bibfnamefont{I.}~\bibnamefont{Cloet}},
  \bibinfo{author}{\bibfnamefont{W.}~\bibnamefont{Bentz}}, \bibnamefont{and}
  \bibinfo{author}{\bibfnamefont{A.~W.} \bibnamefont{Thomas}},
  \bibinfo{journal}{Phys.Lett.} \textbf{\bibinfo{volume}{B642}},
  \bibinfo{pages}{210} (\bibinfo{year}{2006}), \eprint{nucl-th/0605061}.

\end{thebibliography}

\end{document}